\documentclass[10pt,journal,compsoc]{IEEEtran}
\usepackage{graphicx}
\usepackage{times}
\usepackage{epsfig}
\usepackage{graphicx}
\usepackage{amsmath}
\usepackage{amssymb}
\usepackage{overpic}
\usepackage{subfigure}
\usepackage{multirow}
\usepackage{color}
\usepackage{cite}
\usepackage{diagbox}

\usepackage[numbers,sort&compress]{natbib}
\newcommand{\tabincell}[2]{\begin{tabular}{@{}#1@{}}#2\end{tabular}}

\usepackage{color,soul}

\renewcommand{\raggedright}{\leftskip=0pt \rightskip=0pt plus 0cm}

\definecolor{darkcyan}{rgb}{0.0, 0.55, 0.55}
\newcommand{\wh}[1]{{\color{black} #1}}

\begin{document}
\title{Single Image Deraining: From Model-Based to Data-Driven and Beyond}

\author{
	Wenhan~Yang, \textit{Member, IEEE},
	Robby T. Tan, \textit{Member, IEEE},
	Shiqi~Wang, \textit{Member, IEEE}, \\
	Yuming~Fang, \textit{Senior Member, IEEE},
	and~Jiaying~Liu, \textit{Senior Member, IEEE}
	\\
	\IEEEcompsocitemizethanks{
		\IEEEcompsocthanksitem Wenhan Yang and Shiqi Wang are with Department of Computer Science, City University of Hong Kong, Hong Kong, China, (e-mail: \{wyang34, shiqwang\}@cityu.edu.hk).
		\IEEEcompsocthanksitem Robby T. Tan is with Yale-NUS College and the Department of Electrical and Computer Engineering, National University of Singapore, Singapore, (e-mail: tanrobby@gmail.com).
		\IEEEcompsocthanksitem Yuming Fang is with the School of Information Management, Jiangxi University of Finance and Economics, Jiangxi, China, (e-mail: fa0001ng@e.ntu.edu.sg).
		\IEEEcompsocthanksitem Jiaying Liu is with Wangxuan Institute of Computer Technology, Peking University, Beijing, China, (e-mail: liujiaying@pku.edu.cn).
	}
}	

\markboth{IEEE Transactions on Pattern Analysis and Machine Intelligence}
{Shell \MakeLowercase{\textit{et al.}}: Bare Demo of IEEEtran.cls for Computer Society Journals}

\IEEEtitleabstractindextext{
	
\begin{abstract}
\raggedright 
The goal of single-image deraining is to restore the rain-free background scenes of an image degraded by rain streaks and rain accumulation.
The early single-image deraining methods employ a cost function, where various priors are developed to represent the properties of rain and background layers.
Since 2017, single-image deraining methods step into a deep-learning era, and exploit  various types of networks, \textit{i.e.} convolutional neural networks, recurrent neural networks, generative adversarial networks, \textit{etc.}, demonstrating impressive performance.
Given the current rapid development, in this paper, we  provide a comprehensive survey of deraining methods over the last decade.
We summarize the rain appearance models, and discuss two categories of deraining approaches: model-based and data-driven  approaches. 
For the former, we organize the literature based on their basic models and priors. 
For the latter, we discuss developed ideas related to  architectures, constraints, loss functions, and training datasets. 
We present milestones of single-image deraining methods, review a broad selection of previous works in different categories, and provide insights on the historical development route from the model-based to data-driven methods.
We also summarize performance comparisons quantitatively and qualitatively. Beyond discussing the technicality of deraining methods, we also discuss the future directions.
\renewcommand{\baselinestretch}{1}
\end{abstract}

}

\maketitle

\IEEEpeerreviewmaketitle

\IEEEraisesectionheading{
\section{Introduction}\label{sec:introduction}}

\begin{figure}[t]
	\centering
	\subfigure[]{
		\includegraphics[height=4.3cm]{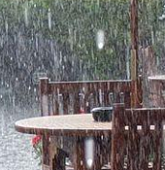}} \hspace{-1mm}	
	\subfigure[]{
		\includegraphics[height=4.3cm]{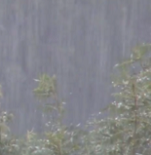}}
	\caption{Different types of visibility degradation caused by rain. 
		\wh{(a) \textit{Rain streaks} cause severe occlusion on the background scene.}
		\wh{(b) \textit{Rain accumulation} significantly degrades the contrast of the scene and reduce the visibility.}
	}
	\vspace{-5mm}
	\label{fig:rain_degradation2}
\end{figure}

Rain introduces visual degradations to captured images and videos.
\textit{Rain streaks} particularly in heavy rain can cause severe occlusion on the background scene.
\textit{Rain accumulation}~\cite{2017_YangRainRemoval}, where distant rain streaks cannot be seen individually and  together with water particles form  a layer of veil on the background, significantly degrades the contrast of the scene and reduce the visibility. Fig.~\ref{fig:rain_degradation2} shows examples of degradation due to rain streaks and rain accumulation.
Human vision and many computer vision algorithms suffer from this degradation, since most of these algorithms assume clear weather, with no interference of rain streaks and rain accumulation. 
Hence, restoring images from rain, called deraining or rain removal, is much desired in many practical applications.

An early study of video deraining was started in 2004 by Garg and Nayar~\cite{garg2004detection}. They analyze rain dynamic appearances, and develop an approach to remove rain streaks from videos.
Kang \textit{et al.}~\cite{ID} was a pioneer in the single image  deraining by publishing a method in 2012.
The method extracts the high-frequency layer of a rain image, and decomposes the layer further into rain and non-rain components using dictionary learning and sparse coding.
Starting from 2017, by the publications of \cite{2017_YangRainRemoval,DetailNet}, data-driven deep-learning methods that  learn features automatically become dominant in the literature.

\begin{figure*}[t]
	\centering
	\subfigure{
		\includegraphics[width=18cm]{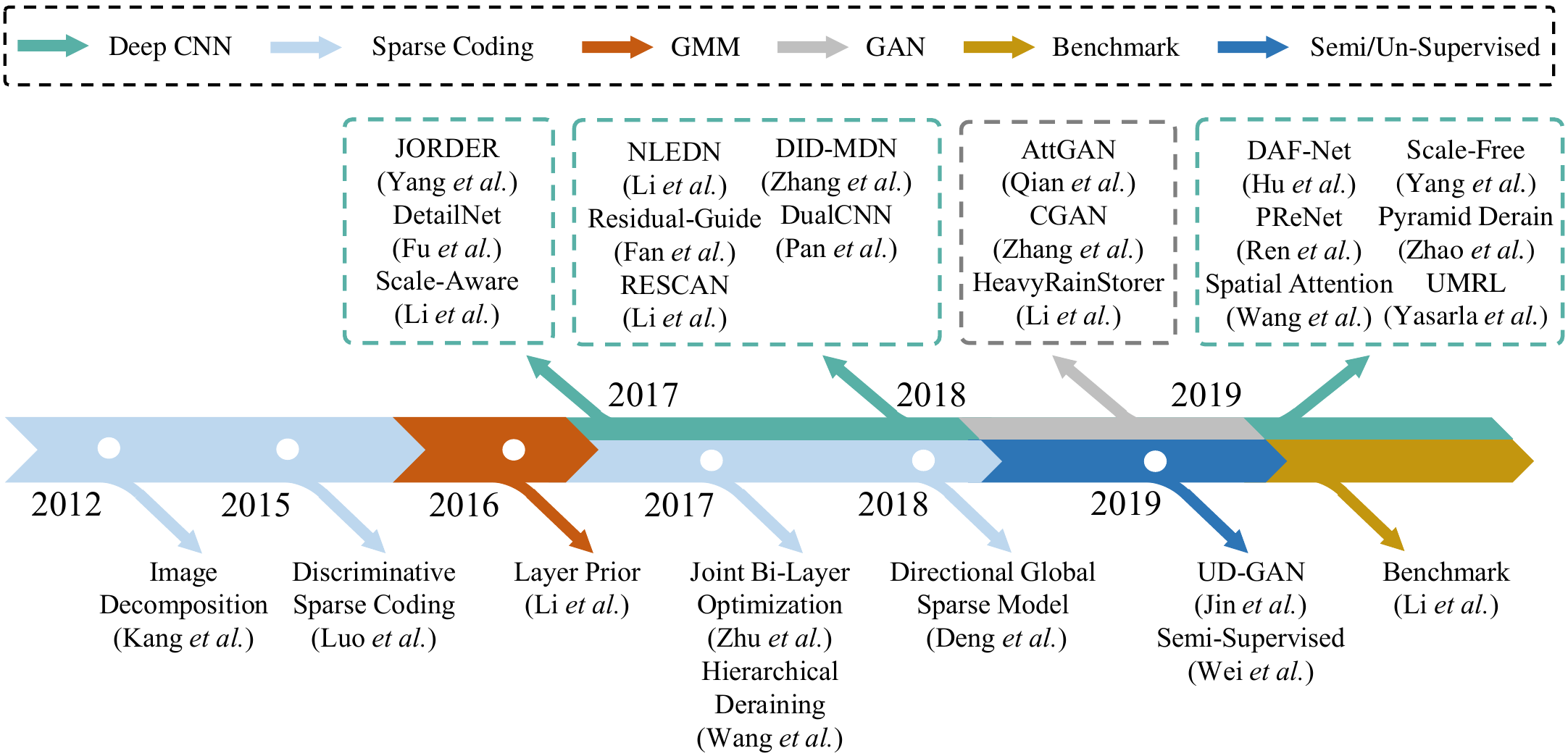}}
	\caption{Milestones of single image deraining methods: image decomposition, sparse coding, Gaussian mixture models, deep convolutional network, generative adversarial network, and semi/unsupervised learning.
	\wh{Before 2017, the typical methods are model-based approach (or non-deep learning approach).
	Since 2017, single-image deraining methods enter into a period  of data-driven approach (or deep learning approach).}
	}
	\label{fig:milestone}
\end{figure*}

In this survey, we focus on single-image deraining, which the aim is to estimate the rain-free background layer of an image degraded by rain streaks and rain accumulation. 
Unlike video deraining methods, which leverage temporal redundancy and dynamics of rain,  single image deraining methods exploit the spatial information of neighboring pixels and the  visual properties of rain and background scenes. 

The milestones of single-image deraining in the past years are
presented in Fig.~\ref{fig:milestone}.
Before 2017, the typical methods are model-based approach (or non-deep learning approach).
The major developments in the model-based approach are driven by the following ideas:
image decomposition (2012), sparse coding (2015), and priors based Gaussian mixture models (2016). 
Since 2017, single-image deraining methods enter into a period  of data-driven approach (or deep learning approach).
The major developments in the data-driven approach are indicated by the following ideas:
deep convolutional network (2017), generative adversarial network (2019), and semi/unsupervised methods (2019).
In 2017-2019, there are more than 30 papers on this deep learning approach, significantly more than the number of deraining papers before 2017.

Model-based methods rely more on the statistical analysis of rain streaks and background scenes.  The methods enforce handcrafted priors on both rain and background layers, then build  a cost function and optimize it.
The priors are extracted from various ways:  Luo \textit{et al.}~\cite{luo2015removing} learn dictionaries for both rain streak and background layers,
Li \textit{et al.}~\cite{li2016rain} build Gaussian mixture models from clean images to model background scenes, and from rain patches of the input image to model rain streaks,
Zhu \textit{et al.}~\cite{Zhu_bilayer} enforce a certain rain direction based on rain-dominated regions so that the background textures can be differentiated from rain streaks.

In recent years, the popularity of data-driven  methods has overtaken model-based methods.
These methods exploits deep networks to automatically extract hierarchical features, enabling them to model more complicated mappings from rain images to clean ones.
Some rain-related constraints are usually injected into the networks to learn more effective features, such as rain masks~\cite{2017_YangRainRemoval}, background features~\cite{Residual_Guide}, \textit{etc}.
Architecture wise, some methods utilize recurrent network~\cite{2017_YangRainRemoval}, or recursive network~\cite{Ren_2019_CVPR} to remove rain progressively.
There are also a series of works focusing on the hierarchical information of deep features, \textit{e.g.}~\cite{NLEDN,PyramidNet}.

While deep networks lead to a rapid progress in deraining performance, many of these deep-learning deraining methods train the networks in a fully supervised way.
This can cause a problem, since to obtain paired images of rain and rain-free images is intractable. The simplest solution is to utilize synthetic images. Yet, there are domain gaps between synthetic rain and real rain images, which can make the deraining performance not optimum.
To overcome the problem, unsupervised/semi-supervised methods that exploit real rain images ~\cite{UD_GAN} and ~\cite{Lil_2019_CVPR} are introduced. 

Our paper aims to provide a comprehensive survey on single-image deraining methods. We believe it can provide a useful starting point to understand the main development of the field, the limitations of existing methods, and the possible future directions.
The rest of the paper is organized as follows.
Section II introduces the rain appearance model.
Section III  provides a detailed survey of single-image rain removal methods, including their synthetic rain models, deraining challenges, methods architectures, and the related technical development.
A particular emphasis is placed on the deep-learning based methods as they offer the most significant progress in the recent years.
Subsequently, Section IV gives detailed discussion on technical developments of network architectures, basic blocks, and summaries of loss functions and databases.
Section V summarizes the quantitative comparisons of a number of single-image rain removal methods and shows qualitative comparisons.
Finally, the paper is concluded in Section VI.

\section{Raindrop Appearance Models}
\label{sec:rain_models}
\begin{figure}[t]
	\centering
	\subfigure{
		\includegraphics[width=7.5cm]{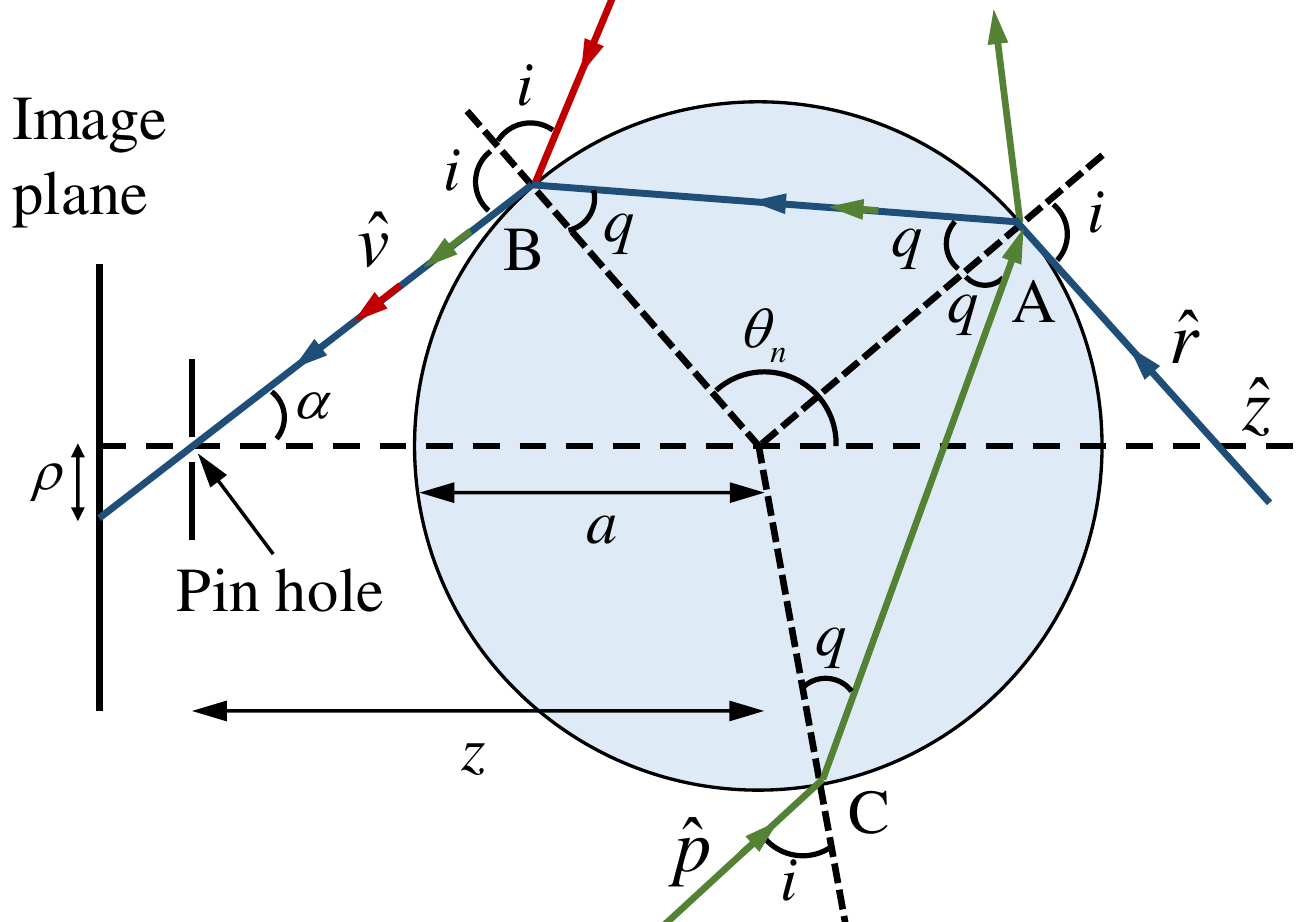}}
	\vspace{-2mm}
	
	\caption{A raindrop's appearance\wh{~\cite{vision_and_rain}} is a complex mapping of the environmental radiance, which is determined by reflection, refraction and internal reflection.}
	\label{fig:raindrop_model}
	\vspace{-3mm}
\end{figure}

%
The shape of a raindrop is usually approximated by a spherical shape~\cite{vision_and_rain}.
As shown in Fig.~\ref{fig:raindrop_model}, considering a point B on the surface of the raindrop
with a surface normal $\hat{n}$, 
rays of light ($\hat{r}$, $\hat{s}$ and $\hat{p}$)  are directed toward the observer via refraction, specular reflection, and internal reflection, respectively.
Hence, the radiance $L(\hat{n})$ at point B is approximated as the 
sum of the radiance $L_r$ of refracted ray, radiance $L_s$
of specularly reflected ray and radiance $L_p$ of internally
reflected ray:
\begin{equation}
\label{eq:eq1}
L(\hat{n}) = L_r(\hat{n}) + L_s(\hat{n}) + L_p(\hat{n}).
\end{equation}
Considering that the radiances depend on the environmental radiance $L_e$ in the direction of the reflected or refracted ray, Eq.~\eqref{eq:eq1} can be expressed as:
\begin{equation}
\label{eqn:raindrop_model}
L(\hat{n}) = R L_e(\hat{r}) + S L_e(\hat{s}) + P L_e(\hat{p}),
\end{equation}
where $R$, $S$ and $P$ denote the fractions of incident environmental
radiance that reaches the camera after refraction,
reflection and internal reflection, respectively. 
We refer to these fractions $(R, S, P)$ as radiance transfer functions.

Moreover, we can reach the composite raindrop model written as~\cite{vision_and_rain}:
\begin{equation}
\label{eqn:raindrop_model2}
L(\hat{n}) = \left(1-k(i,\mu)\right)^2 L_e(\hat{r}) + S L_e(\hat{s}) + P L_e(\hat{p}),
\end{equation}
where $i = (\pi − \theta n + \alpha)$ is the incident angle, $\mu$ is the
refractive index of the water and $k$ is the Fresnel’s reflectivity 
coefficient for unpolarized light.
Based on the statistics from~\cite{vision_and_rain}, the radiance of the
raindrop is mainly decided by the refraction, and the appearance of
the raindrop is mainly based on refraction through the
drop.

%

For a moving raindrop, its appearance changes significantly.
The raindrop becomes a rain streak, and its appearance relies on the brightness of the raindrop, background scene radiances, and camera's exposure time. The change of pixel's intensity value caused by a rain streak can be approximated as~\cite{vision_and_rain}:
\begin{align}
\Delta I &= -\beta I_b	+ \alpha,
\label{eqn:rain_streaks}
\end{align}
with $\beta = \frac{\tau}{T}$, and  
$\alpha = \tau \overline{E}_r$, where 
$\tau$ is the time when a drop remains within a pixel, 
and $T$ is the exposure time. 
$\overline{E}_r$ is the time-averaged irradiance caused by the drop.
Based on Eq.~\eqref{eqn:rain_streaks}, we can reach two conclusions: 
1) A raindrop causes an intensity change and
moves faster than the integration
time of a typical video camera;
2) the intensity change of a rain streak correlates linearly to the background intensity $I_b$.
Based on the derived numerical bounds~\cite{vision_and_rain}, empirically we obtain: $0 < \beta < 0.039$ and $0 < \tau < 1.18$.
In most real cases, $\alpha$ dominates the appearance of $\Delta I$, thus: 
\begin{eqnarray}
\Delta I = \alpha.
\label{eqn:app_simple}
\end{eqnarray}
As a result, in most rain synthetic models, \textit{rain streaks are assumed to be superimposed on the background image}.

\begin{table*}[t]
	\centering
	\footnotesize
	\caption{Summary of rain synthetic models in the literature.}
	\label{tab:summary_rain_models}
	\begin{tabular}{cccc}
		\hline
		Method                     & Degradation Factors                   & Main Features                                           & Publication \\
		\hline
		\hline
		Additive Composite Model (ACM)   & Streak                                & Simple and effective                                    & Li \textit{et al.} 2016~\cite{li2016rain} \\ 
		Screen Blend Model (SBM)   & Streak                                & Streaks and backgrounds are combined nonlinearly        & Luo \textit{et al.} 2015~\cite{luo2015removing} \\ 
		Heavy Rain Model (HRM)     & Streak, Accumulation                  & Overlapping streaks generating accumulation             & Yang \textit{et al.} 2017~\cite{2017_YangRainRemoval} \\
		Rain Model with Occlusion (ROM) & Streak, Occlusion                     & Considering rain occlusions                             &  Liu \textit{et al.} 2018~\cite{J4R} \\ 
		Comprehensive Rain Removal (CRM) & \tabincell{c}{Streak, Occlusion, \\ Accumulation, Flow} & Considering comprehensive visual degradation            &  Yang \textit{et al.} 2019~\cite{Yang_2019_CVPR} \\ 
		Depth-Aware Rain Model (DARM) & Streak, Accumulation                  & Streaks and accumlation modeling correlated with depth & Hu \textit{et al.} 2019~\cite{Hu_2019_CVPR}  \\
		\hline
	\end{tabular}
\end{table*}

\section{Literature Survey}
\label{sec:survey}

In this section, we first review a few rain synthesis models proposed in some existing methods. Unlike in the previous section (Sec.~\ref{sec:rain_models}), the models we discuss here are only loosely based on physics and thus, to our knowledge, their correctness has not been verified both theoretically or experimentally. Despite this, the methods that use these models show, to some extent, the effectiveness of the models on real-image deraining. Having discussed various rain synthetic models, we briefly explain the challenges in image deraining, and  subsequently survey on  existing deraining methods comprehensively.

\subsection{Synthetic Rain Models}

\vspace{0.3cm}

\noindent \textbf{Additive Composite Model}
The most simple and popular rain model used in existing studies is the additive composite model~\cite{li2016rain,ID}, which  follows Eq.~\eqref{eqn:app_simple} and is expressed as:
\begin{equation}
\label{eq:additive_composite_model}
\mathbf{O} = \mathbf{B}+\mathbf{S},
\end{equation}
where $\mathbf{B}$ denotes the background layer, and $\mathbf{S}$ is the rain streak layer. 
$\mathbf{O}$ is the image degraded by rain streaks. Here, the model assumes that the appearance of rain streaks is simply superimposed to the background, and there is no rain accumulation in the rain degraded image.

\vspace{0.4cm}

\noindent \textbf{Screen Blend Model}
Luo \textit{et al.}~\cite{luo2015removing} propose a non-linear composite model, called screen-blend model:
\begin{equation}
\label{eq:screen_blend_model}
\mathbf{O} = 1 - \left(1-\mathbf{B}\right) \circ \left(1-\mathbf{S}\right)=\mathbf{B}+\mathbf{S}-\mathbf{B} \circ\mathbf{S},
\end{equation}
where $\circ$ denotes the operation of point-wise multiplication. 
Unlike the additive composite model in Eq.~(\ref{eq:additive_composite_model}), the background and rain layers influence the appearance of each other.
Luo \textit{et al.}~\cite{luo2015removing} claim that the screen blend model can model some visual properties of real rain images, such as the effect of internal reflections, and thus generate visually more authentic rain images.
The combination of rain and background layers are signal-dependent. Implying, when the background is dim, the rain layer will dominate the appearance of the rain image; and, when the background is bright, the background layer will dominate the image.

\vspace{0.4cm}

\noindent \textbf{Heavy Rain Model}
Yang \textit{et al.}~\cite{2017_YangRainRemoval} propose a rain model that includes both rain streaks and rain accumulation. This is the first model in the deraining literature that includes the two rain phenomena.
Rain accumulation or rain veiling effect is a result of water particles in the atmosphere and distant rain-streaks that cannot be seen individually.
The visual effect of rain accumulation is similar to mist or fog, which leads to low contrast.
Considering two main aspects of rain: the Koschmieder model to approximate the visual appearance of a scene in a turbid medium, and overlapping rain streaks that have different directions and shapes, a novel rain model is introduced:
\begin{equation}
\label{eq:heavy_rain_model}
\textbf{O} = \alpha \circ \left(\mathbf{B}+\sum_{t=1}^{s}\mathbf{S}_t\right)+(1-\alpha)\mathbf{A},
\end{equation}
where $\textbf{S}_t$ denotes the rain-streak layer that has the same streak direction.
$t$ indexes the rain-streak layer and $s$ is the maximum number of the rain-streak layers.
$\textbf{A}$ is the global atmospheric light, and $\alpha$ is the atmospheric transmission.

\vspace{0.4cm}

\noindent \textbf{Rain Model with Occlusion}
Liu \textit{et al.}~\cite{J4R} extend the heavy rain model to an occlusion-aware rain model for modeling rain in video.
The model separates rain streaks into two types: transparent rain streaks that are added to the background layers, and opaque rain streaks that totally occlude the background layers.
The locations of these opaque rain streaks are indicated by a map, called the reliance map.
The formulation of this rain model is expressed as: 
\begin{equation}
\label{eq:occlusion_rain_model}
\mathbf{O} = \beta \circ \left(\mathbf{B}+\sum_{t=1}^{s}\mathbf{S}_t\right) + (1-\beta) \circ \mathbf{R},
\end{equation} 
where $\textbf{R}$ is the rain reliance map and $\beta$ is  defined as:
\begin{equation}
\label{eq:rain_alliance}
\mathbf{\beta}_t\left(i,j\right) = \left\{
\begin{matrix}
1, & \text{ if } (i,j)\in \Omega_{\mathbf{S}},\\
0, & \text{ if } (i,j)\notin \Omega_{\mathbf{S}},\\
\end{matrix}
\right.
\end{equation}
where $\Omega_{\mathbf{S}}$ is defined as the rain occluded  region.

\vspace{0.4cm}

\noindent \textbf{Comprehensive Rain Model}
Yang \textit{et al.}~\cite{Yang_2019_CVPR} combine all above mentioned degradation factors into a comprehensive rain model for modeling rain appearance in video. 
It considers the temporal properties of rain scenes, particularly the fast-changing rain accumulations that usually cause flicker. This visible intensity changes along the temporal dimension is called rain accumulation flow.
Besides, it also considers other factors including rain streaks, rain accumulation, and rain occlusion, which are formulated as:
\begin{equation}
\label{eq:rain_gen_accu}
\mathbf{O} = \beta \circ \left[ \left(\mathbf{B}+\sum_{t=1}^{s}\mathbf{S}_t\right) + (1-\alpha)\mathbf{A} + \mathbf{U} \right] + (1-\beta) \circ \mathbf{R},
\end{equation}
where $\mathbf{U}$ is the rain accumulation flow. 

\vspace{0.4cm}

\noindent \textbf{Depth-Aware Rain Model}
Hu \textit{et al.}~\cite{Hu_2019_CVPR} further connect $\alpha$ to the scene depth $d$, to create a depth-aware rain model:
\begin{equation}
\label{eq:depth_aware_model}
\mathbf{O} = \mathbf{\beta} \circ \left(1- \sum_{t=1}^{s}\mathbf{S}_t - (1-\alpha) \mathbf{A} \right) + \sum_{t=1}^{s}\mathbf{S}_t + \alpha \mathbf{A},
\end{equation}
where $\mathbf{S}_t$ and $\alpha$ are connected with the scene depth written as:
\begin{align}
\label{eq:distance_model}
\mathbf{S}_t(i, j) & = \mathbf{S}_{\text{Pattern}}(i, j) \cdot t_r(i, j), \nonumber \\
t_r(i, j) &= e^{-\alpha \text{max}(d_\text{M}, d(i,j))}, \\
A(i,j) &= 1-e^{-\beta  d(i,j)}, \nonumber 
\end{align}
where $\mathbf{S}_{\text{Pattern}}(i, j)$ is an intensity image of uniformly-distributed rain streaks in the image space, and $t_r(i,j)$ is the rain streak intensity map relying on the depth. $d(i,j)$ denotes the depth and $\alpha$ controls the rain streak intensity. $\beta$ determines the thickness of fog, where a larger $\beta$ denotes a thicker fog.

\vspace{0.4cm}

\noindent \textbf{Discussions} Following these different rain models, various rain degradation can be synthetically rendered.
\wh{A summary of rain synthetic models in the literature is provided in Table~\ref{tab:summary_rain_models}.}
In general, heavy rain models~\cite{2017_YangRainRemoval,Lil_2019_CVPR} and depth-aware rain models~\cite{Hu_2019_CVPR} cover the most comprehensive rain degradation for single rain image synthesis.
However, as we mentioned in the beginning of this section that all these models are heuristic; implying that they might not entirely correct physically. Nevertheless, as shown in the literature, they can be effective, at least  to some extent, for image deraining.

\subsection{Deraining Challenges}
The goal of single image deraining is to recover the clean and rain-free background scene from a rain degraded image. 
However, there are a few challenges to accomplish the goal:
\begin{itemize}
	\item 
	\textbf{Difficulties in modeling rain images}
	In the real world,  rain can visually appear in many different ways.
	Rain streaks can vary in terms of  sizes, shapes, scales, densities, directions, etc. Similarly, rain accumulation depends on various water-particles and  atmospheric conditions. Moreover, rain appearance significantly relies also on the textures and the depth of the background scenes.
	All these cause difficulties in modeling the appearance of rain, which consequently cause   the rendering of physically-correct rain images to be a complex task. 
	
	\item \textbf{Ill-posedness of deraining problem}
	Even with a simple rain model that considers only rain-streaks, to estimate the background scene from a degraded image is an ill-posed problem. 
	The reason is that we only have the pixel intensity values produced by lights carrying fused information of rain and background scenes. To make the matter worse, in some cases the background information can be totally occluded by rain streaks or dense rain accumulation or both.
	
	\item
	\textbf{Difficulties in finding proper priors}
	As rain and background information might overlap in the feature space, it is non-trivial to separate them. Background textures can be falsely deemed as rain, resulting in incorrect deraining. Hence, strong priors for background textures and rain are necessary. However, finding these priors is difficult, since background textures are diverse, and some have similarity to the appearance of rain-streaks or rain accumulation.
	
	\item 
	\textbf{Real paired ground-truths}
	Most of deep-learning methods rely on paired rain and clean background images to train their networks. However, to obtain real rain images and their exact pairs of clean background images is intractable. 
	Even for a static background, lighting conditions always change. This difficulty does not only impact on deep-learning methods, but also for evaluating the effectiveness of any method. Currently, for qualitative evaluation, all methods rely on human subjective judgment on whether the restored images are good; and for quantitative evaluation, all current methods rely on synthetic images. Unfortunately, up to now, there are significant gaps between synthetic and real images.
\end{itemize}
In the following section, we will discuss how existing deraining methods deal with these challenges.

\begin{figure*}[htbp]
	\centering
	\subfigure{
		\includegraphics[width=18cm]{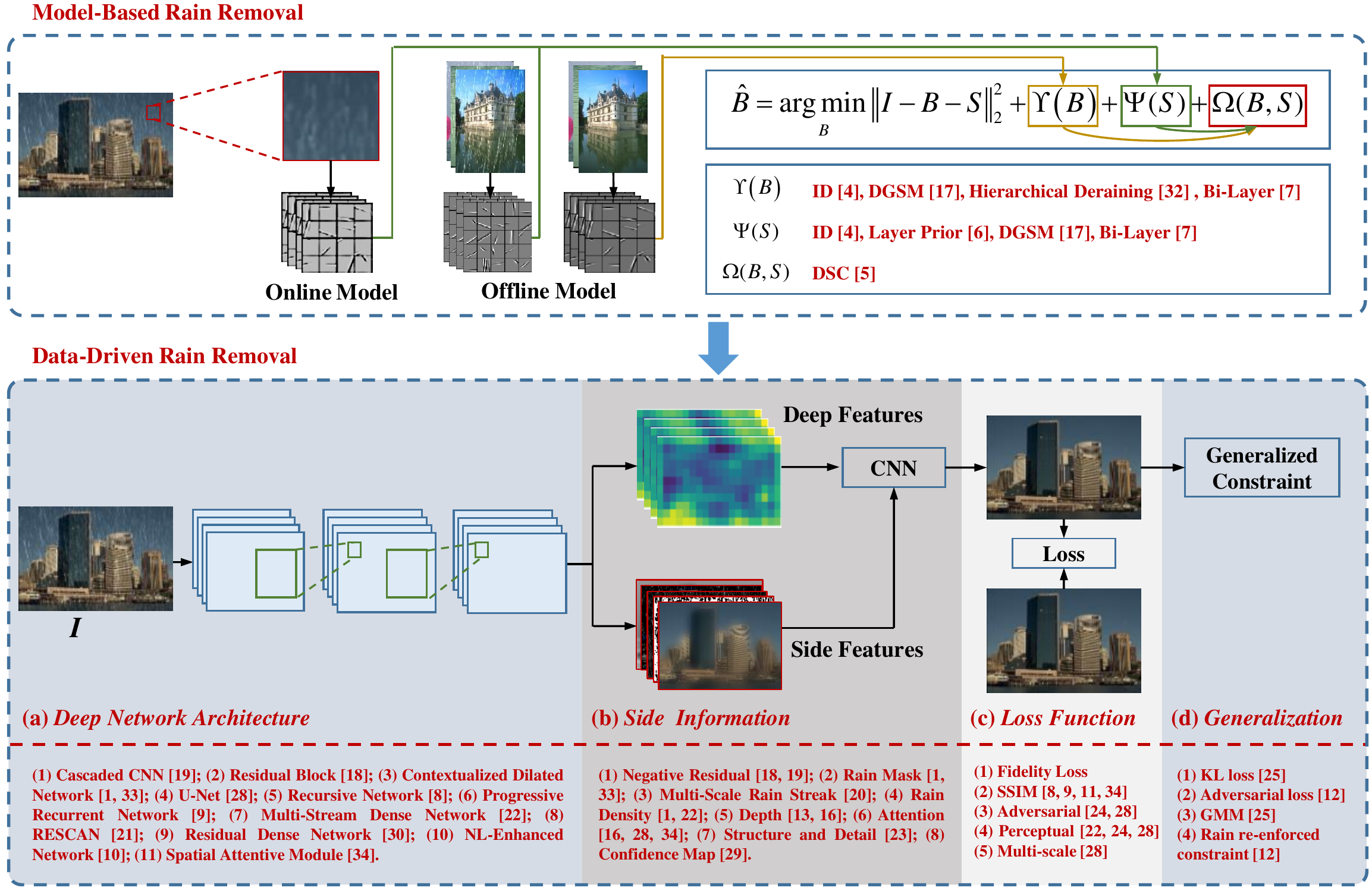}}
	\caption{The improvement of single-image rain removal, from model-based to data-driven approaches.
	\wh{The model-based methods employ optimization frameworks for deraining. They rely on the statistical analysis of rain streaks and background scenes, and enforce handcrafted priors on both rain and background layers.}
	\wh{
	Data-driven approaches exploit deep networks to automatically extract hierarchical features, enabling them to model more complicated mappings from rain images to clean ones.
	Some rain-related constraints are  injected into the networks to learn more effective features.}
	}
	\label{fig:overall}
\end{figure*}

\subsection{Single-image Deraining Methods}

We categorize  single-image deraining methods into two basic approaches: \textit{model-based} (non-deep-learning) and \textit{data-driven} (deep-learning) approaches.
We will discuss the existing methods of the two approaches in detail in the subsequent sections. \wh{A summary of previous works is given in Table~\ref{tab:summary}}.
In addition, for the sake of completeness, we will also briefly discuss adherent raindrop removal methods, since adherent raindrops (i.e., water droplets attached to a lens or windscreen) are also part of rain degradation; although, in some situations, they can be avoided by placing the camera under a shelter.

\begin{table*}[htbp]
	\scriptsize
	\caption{An overview of single-image rain removal methods.}
	\label{tab:summary}
	\centering
	\begin{tabular}{c|c|c|c|l|c}
		\hline
		Method                      & Category                        & Rain Model & \begin{tabular}[c]{@{}c@{}} Variables or Priors \end{tabular} & \qquad \qquad \qquad \qquad \qquad \quad Key Idea & Publication                         \\
		\hline
		\hline
		Image Decomposition         & \multirow{16}{*}{\tabincell{c}{Sparse \\ Representation}}  &  ACM       &   \tabincell{c}{Low/high-frequency; \\ Rain/No-rain dictionaries.}   & 
		\tabincell{l}{Rain images are decomposed into
			the low/high-frequency~(HF) \\  parts. Then, the ``nonrain component'' is removed from HF part \\ by dictionary learning and sparse coding. 
			details.}  
		&  Kang \textit{et al.} 2012~\cite{ID} \\ 
		\cline{1-1}\cline{3-6}
		DSC                         &                                 &  SBM       &    \tabincell{c}{Rain/No-rain codes; \\ Rain/No-rain dictionaries.}  &
		\tabincell{l}{Patches of two layers are sparsely approximated by very high \\  discriminative codes over a learned dictionary with strong \\  mutual exclusivity property.}
		& Luo \textit{et al.} 2015~\cite{luo2015removing} \\ 
		\cline{1-1}\cline{3-6}
		Bi-Layer Optimization       &                                 &  ACM      &     \tabincell{c}{Centralized sparse prior; \\ Rain direction prior; \\ Rain layer prior.}  &         
		\tabincell{l}{A joint optimization process is used that alternates between \\ removing rain-streak details from the estimated background and \\ removing non-streak details from the estimated rain streak layer. }		
		& Zhu \textit{et al.} 2017~\cite{Zhu_bilayer}                           \\
		\cline{1-1}\cline{3-6}
		Hierarchical Deraining       &                                 &  ACM      &     \tabincell{c}{High/low-frequency; \\ Sensitivity of variance \\ across color channels; \\  Principal direction of \\ an image patch. }  &         
		\tabincell{l}{A three-layer hierarchical scheme is designed. The first layer \\ uses sparse coding to classify the high-frequency part into \\ rain/snow  and non-rain/snow components. The second layer \\ relies on guided filtering. The third layer enhances the visual \\ quality with the sensitivity of variance across color channels. }		
		& Wang \textit{et al.} 2017~\cite{Hierarchical_deraining}                           \\		
		\cline{1-1}\cline{3-6}
		DGSM                        &                                 &  ACM       &    \tabincell{c}{Unidirectional TV; \\ Rotation Angle.}  & 
		\tabincell{l}{A global sparse model is formulated to consider the intrinsic \\ 
			directional, structural knowledge of rain streaks, and the \\ property of image back-ground information.}
		& Deng \textit{et al.} 2018~\cite{DENG2018662} \\ 
		\hline
		LP                          & GMM                             &  ACM       &    \tabincell{c}{Gaussian mixture model; \\ Total variation.}  & 
		\tabincell{l}{Two patch-based priors for the background and rain layers \\ are built based on Gaussian mixture models and can \\ accommodate multiple orientations and scales of rain streaks.} & Li \textit{et al.} 2016~\cite{li2016rain} \\ 
		\hline
		CNN                         & \multirow{40}{*}{Deep CNN}      &  \wh{Raindrop}       & Clean image. &  \tabincell{l}{A three-layer CNN is constructed to learn the mapping between \\ 
			the corrupted image patches to clean ones.} & Eigen \textit{et al.} 2013~\cite{Eigen_2013_ICCV} \\
		\cline{1-1}\cline{3-6}
		\tabincell{c}{JORDER\\JORDER-E}                      &                                 &  HRM       &   
		\tabincell{c}{Binary rain mask; \\  Rain intensity; \\ Residual.} &  
		\tabincell{l}{A multi-task architecture is constructed to learn the binary rain \\ streak map, the appearance of rain streaks, and the clean \\ background. A recurrent network is also built to remove rain \\ streaks and clears up the rain accumulation progressively.} & \tabincell{l}{Yang \textit{et al.} 2017~\cite{2017_YangRainRemoval} \\ Yang \textit{et al.} 2019~\cite{Yang_2019}}     \\ 
		\cline{1-1}\cline{3-6}
		DetailNet                   &                                 &  ACM       &  Residual.      &   \tabincell{l}{A deep detail network taking as input the high frequency detail \\ and predict the residual between the input image and the ground \\ truth image.} & \tabincell{l}{Fu \textit{et al.} 2017~\cite{DetailNet} \\ Fu \textit{et al.} 2017~\cite{DetailNet2} } \\ 
		\cline{1-1}\cline{3-6}
		
		Scale-Aware                 &                                 &  HRM       &  \tabincell{c}{Rain streak; \\ Transmission map.}  &
		\tabincell{l}{Parallel sub-networks are built to predict different scales of rain \\ streaks and the transmission maps.}
		& Li \textit{et al.} 2017~\cite{Li_scale} \\ 
		\cline{1-1}\cline{3-6}
		NLEDN                       &                                 &  ACM       &  Residual.   & 
		\tabincell{l}{An auto-encoder network is built with non-locally enhanced  \\ 
			dense blocks, where a non-local feature map weighting follows \\ 
			four densely connected convolution layers.
		}
		& Li \textit{et al.} 2018~\cite{NLEDN} \\ 
		\cline{1-1}\cline{3-6}
		Residual-Guide              &                                 &  ACM       &  Residual.  &  
		\tabincell{l}{The residuals generated from shallower blocks are used to guide \\ 
			deeper blocks. The negative residual is predicted coarse to fine \\
			and the outputs of different blocks are fused finally.}  &  Fan \textit{et al.} 2018~\cite{Residual_Guide} \\ 
		\cline{1-1}\cline{3-6}	
		RESCAN                      &                                 &  ACM       &  \tabincell{c}{Residual; \\ Intermediate results.} & \tabincell{l}{A recurrent network is constructed and the rain removal result \\  of the previous stage is fed into the next stage. The information \\ is flowed across stages at the feature level.}  &  Li \textit{et al.} 2018~\cite{li2018recurrent} \\ 
		\cline{1-1}\cline{3-6}			
		DID-MDN                     &                                 &  ACM       &  \tabincell{c}{Rain density; \\ Residual.}  & \tabincell{l}{A multi-path densely connected network is constructed to \\ automatically detect the rain-density to guide the rain removal.}  & Zhang \textit{et al.} 2018~\cite{DID-MDI} \\ 
		\cline{1-1}\cline{3-6}			
		DualCNN                     &                                 &  ACM       &  \tabincell{c}{Structure layer; \\ Detail layer.} &  \tabincell{l}{A dual CNN estimates the two parts of the target signal: \\ structures and details for a series of low-level vision tasks.}  & Pan \textit{et al.} 2018~\cite{Pan_2018_CVPR} \\ 
		\cline{1-1}\cline{3-6}			
		DAF-Net                     &                                 &  DARM      & \tabincell{c}{Depth; \\ Attention; \\ Residual. }  &  
		\tabincell{l}{The work creates a \textit{RainCityscapes} dataset related to the scene \\ 
			depth. A deep network is developed based on a depth-guided \\ attention mechanism to predict the residual map.}   & Hu \textit{et al.} 2019~\cite{Hu_2019_CVPR} \\ 
		\cline{1-1}\cline{3-6}		
		\tabincell{c}{Spatial Attention + \\ Dataset} &               &  ACM       &  \tabincell{c}{Spatial attention; \\ Residual.} &  \tabincell{l}{A high-quality dataset including 29.5K paired real rain and \\ pseudo real non-rain images is constructed. Then, a novel \\ spatial attentive network is built to effectively learn \\ discriminative features  for rain removal from local to global.}   & Wang \textit{et al.} 2019~\cite{Wang_2019_CVPR} \\ 
		\cline{1-1}\cline{3-6}		
		PReNet                      &                                 &  ACM       &  \tabincell{c}{Residual; \\ Intermediate results.}  &   \tabincell{l}{The network utilizes recursive computation at two levels:  \\
			1) Progressive ResNet is built by repeatedly unfolding a \\ shallow ResNet; 2) The ResNet performs stage-wise operations \\ processing the input and intermediate results progressively. }       & Ren \textit{et al.} 2019~\cite{Ren_2019_CVPR} \\ 
		\cline{1-1}\cline{3-6}		
		Scale-Free                  &                                 &  HRM       &  \tabincell{c}{Residual; \\ Wavelet decomposition \\ results. }  &   \tabincell{l}{A recurrent network performs two-stage deraining: 1) rain \\ 
			removal on the low-frequency component; 2) recurrent detail \\
			recovery on high-frequency components guided by the \\
			recovered low-frequency component.}  & Yang \textit{et al.} 2019~\cite{Scale_free} \\ 
		\cline{1-1}\cline{3-6}		
		PyramidDerain               &                                 &  ACM       & \tabincell{c}{Gaussian-Laplacian \\ pyramid.} & \tabincell{l}{The proposed network combines Gaussian Laplacian image \\ pyramid decomposition and the deep neural network. Recursive \\ and residual network structures are employed to aggregate the \\ features at different layers. } & Fu \textit{et al.} 2019~\cite{PyramidNet} \\ 
		\cline{1-1}\cline{3-6}		
		UMRL               &                                 &  ACM       & \tabincell{c}{Confident map.} & \tabincell{l}{The network is guided based on the confidence measure about \\ the estimate. The cycle spinning is introduced to remove \\ artifacts and improve the deraining performance. } & Yasarla \textit{et al.} 2019~\cite{Yasarla_2019_CVPR} \\ 
		\hline		
		AttGAN                      &      \multirow{8}{*}{GAN}       &  \wh{Raindrop} &  \tabincell{c}{Attention map; \\ Clean image.}     &   \tabincell{l}{Visual attention is injected into both the generative and \\
			discriminative networks for learning to attend raindrop regions \\ and percept their surroundings.}  & Qian \textit{et al.} 2018~\cite{Qian_2018_CVPR} \\ 
		\cline{1-1}\cline{3-6}
		CGAN                        &                                 &  ACM      & Clean image.  &  \tabincell{l}{The work directly applies a multi-scale conditional generative \\ adversarial network to address single image de-raining task.}  &  Zhang \textit{et al.} 2019~\cite{CGAN_rain} \\ 
		\cline{1-1}\cline{3-6}		
		HeavyRainRestorer           &                                 &   HRM     & \tabincell{c}{Transmission map; \\ Atmospheric light; \\ Rain streak; \\ Clean image.} & \tabincell{l}{A two-stage network is built: a physics-driven model followed \\ by a depth-guided generative adversarial refinement.}   & Li \textit{et al.} 2019~\cite{Lil_2019_CVPR} \\ 
		\hline
		Semi-supervised CNN         & \multirow{4}{*}{\tabincell{c}{Semi/Un- \\ Supervised}}  &  ACM   & \tabincell{c}{Rain streak; \\ Residual. } &  \tabincell{l}{A semi-supervised learning method formulates the residual as a \\ specific parametrized rain streak distribution between an input \\ 
			rainy image and its expected network output. }  & Wei \textit{et al.} 2019~\cite{Wei_2019_CVPR} \\ 
		\cline{1-1}\cline{3-6}	
		UD-GAN                      &                                 &   ACM      & \tabincell{c}{Rain streak; \\ Residual. }  &  \tabincell{l}{ An Unsupervised Deraining Generative Adversarial Network is \\ built to introduce self-supervised constraints, the intrinsic priors \\ extracted from unpaired rainy and clean images.}   & Jin \textit{et al.} 2019~\cite{UD_GAN}        \\
		\hline
		Benchmark                   &          Benchmark              & \tabincell{c}{ ACM + \\  \wh{Raindrop}}  &  --  & \tabincell{l}{It provides extensive study and evaluation of existing single \\ image 
			deraining algorithms with a new proposed large-scale \\ dataset  
			including both synthetic and real-world rainy images of \\ various rain types.} & Li \textit{et al.} 2019~\cite{Li_2019_CVPR} \\
		\hline                   
	\end{tabular}
\end{table*}

\subsubsection{Model-based Methods}

Existing model-based methods employ optimization frameworks for deraining, as shown in the top panel of Fig.~\ref{fig:overall}. These methods deal only with rain streaks and ignore the presence of rain accumulation. A general optimization framework can be expressed as:
\begin{equation}
	\label{eq:map}
	\hat{\mathbf{B}} = \arg\min_{\mathbf{B}} \|\mathbf{I}-\mathbf{B}-\mathbf{S}\|_2^2 + \Upsilon\left(\mathbf{B}\right)+\Psi\left(\mathbf{S}\right) + \Omega\left(\mathbf{B}, \mathbf{S}\right),
\end{equation}
where $\Upsilon\left(\mathbf{B}\right)$ denotes the priors on the background layers, $\Psi\left(\mathbf{S}\right)$ represents the priors on rain streak layers, and $\Omega\left(\mathbf{B}, \mathbf{S}\right)$ is the joint prior to describe the intrinsic relationship between rain streaks and background layers. 
Different prior terms are designed to better describe and separate the rain streak from the background layers.

\vspace{0.4cm}

\noindent \textbf{Sparse Coding Methods}
Sparse coding~\cite{M_Elad} represents the input vectors as a sparse linear combination of basis vectors.
The collection of these basis vectors is called  dictionary, which is used to reconstruct the certain type of signals, \textit{e.g.} rain streaks and background signals in the deraining problem.
Lin \textit{et al.}~\cite{ID} make the first attempt on single-image deraining via \textit{image decomposition} using a   morphological component analysis. The initially extracted high-frequency component of a rain image is further decomposed into rain and non-rain components by dictionary learning and sparse coding.
This pioneer work successfully removes sparse light rain streaks.
However, it significantly relies on the bilateral filter preprocessing, and thus generates blurred background details.

In a successive work, Luo~\textit{et al.}~\cite{luo2015removing} enforce the sparsity of rain, and introduce a mutual exclusivity property into a discriminative sparse coding (DSC) to facilitate accurately the separation of the rain/background layers from their non-linear composite.
Benefiting from the mutual exclusivity property, the DSC preserves clean texture details; however, it shows some  residual rain streaks in the output, particularly for large and dense rain streaks.
To further improve the modeling capacity, Zhu \textit{et al.}~\cite{Zhu_bilayer} construct an iterative layer separation process to remove rain streaks from the background layer, as well as to remove background's texture details from the rain streak layer using layer-specific priors.
Quantitatively, the method obtains comparable performance on some synthetic datasets with that of deep learning-based methods published in the same period of time, \textit{i.e.} JORDER~\cite{2017_YangRainRemoval} and DDN~\cite{DetailNet}.
However, qualitatively on real images, the method tends to fail in handling heavy rain cases, where rain streaks may move in different directions.

To model rain streak directions and sparsity, Deng \textit{et al.}~\cite{DENG2018662} formulate a directional group sparse model (DGSM), which includes three sparse terms representing the intrinsic directional and structural knowledge of rain streaks.
It can effectively remove blurred rain streaks but fail to remove sharp rain streaks.

\vspace{0.4cm}

\noindent \textbf{Gaussian Mixture Model} 
Li \textit{et al.}~\cite{li2016rain} apply Gaussian mixture models (GMMs) to model both rain and background layers. 
The GMMs of the background layer is obtained off-line from  real images with diverse background scenes.
A selected rain patch from the input image that has no background textures is proposed to train the GMMs of the rain layer.
The total variation is utilized to remove small sparkle rains. 
The method is capable to effectively remove rain streaks of small and moderate scales, but fail to handle large and sharp rain streaks.

\subsubsection{Deep Learning Based Methods}

\vspace{0.2cm}

\noindent \textbf{Deep CNNs}
The era of deep-learning deraining starts in year 2017.
Yang \textit{et al.}~\cite{2017_YangRainRemoval} construct a joint rain detection and removal network. It can handle  heavy rain, overlapping rain streaks, and rain  accumulation.
The network can detect rain locations by predicting the binary rain mask, and take a recurrent framework to remove rain streaks and clear up rain accumulation progressively. 
The method achieves good results in heavy rain cases.
However, it might falsely remove vertical textures and generate underexposed illumination.

In the same year, Fu \textit{et al.}~\cite{DetailNet,DetailNet2} made an attempt to remove rain streaks via a deep detail network~(DetailNet). The network takes only the high frequency details as input, and predicts the residue of the rain and clean images.
The paper shows that  removing the background information in the network input is beneficial, as doing so makes the training easier and more stable.
However, the method still cannot handle large and sharp rain streaks.

Following Yang \textit{et al.}~\cite{2017_YangRainRemoval} and Fu \textit{et al.}~\cite{DetailNet,DetailNet2}, many CNN based methods~\cite{Li_scale, NLEDN, Residual_Guide, li2018recurrent, DID-MDI, Pan_2018_CVPR} are proposed.
These methods employ more advanced network architectures and  injecting new rain related priors.
They achieve better results both quantitatively and qualitatively.
However, due to the limitation of their fully supervised learning paradigm, namely using synthetic rain images, they tend to fail when dealing with some conditions of real rain that has never been seen during training.

\begin{figure}[t]
	\centering
	\subfigure{
		\includegraphics[width=8.7cm]{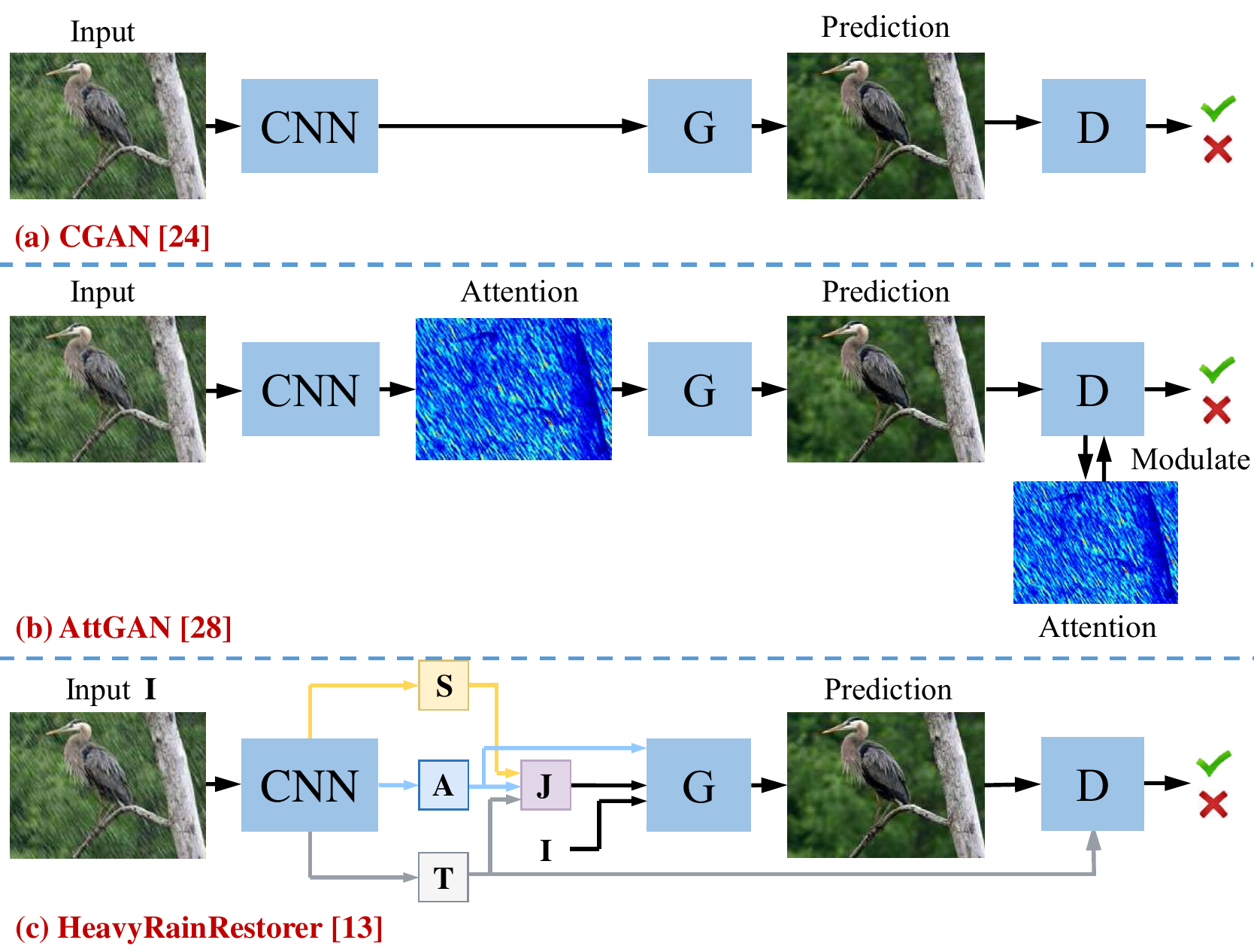}}
	\vspace{-4mm}
	\caption{Summary of GAN-based rain removal methods.
	\wh{To capture some visual properties of rain that cannot be modeled and synthesized, the adversarial learning is introduced to 	reduce the domain gaps between the generated results  and real clean images.}}
	\vspace{-2mm}	
	\label{fig:GAN}
\end{figure}

\begin{figure}[t]
	\centering
	\subfigure{
		\includegraphics[width=8.7cm]{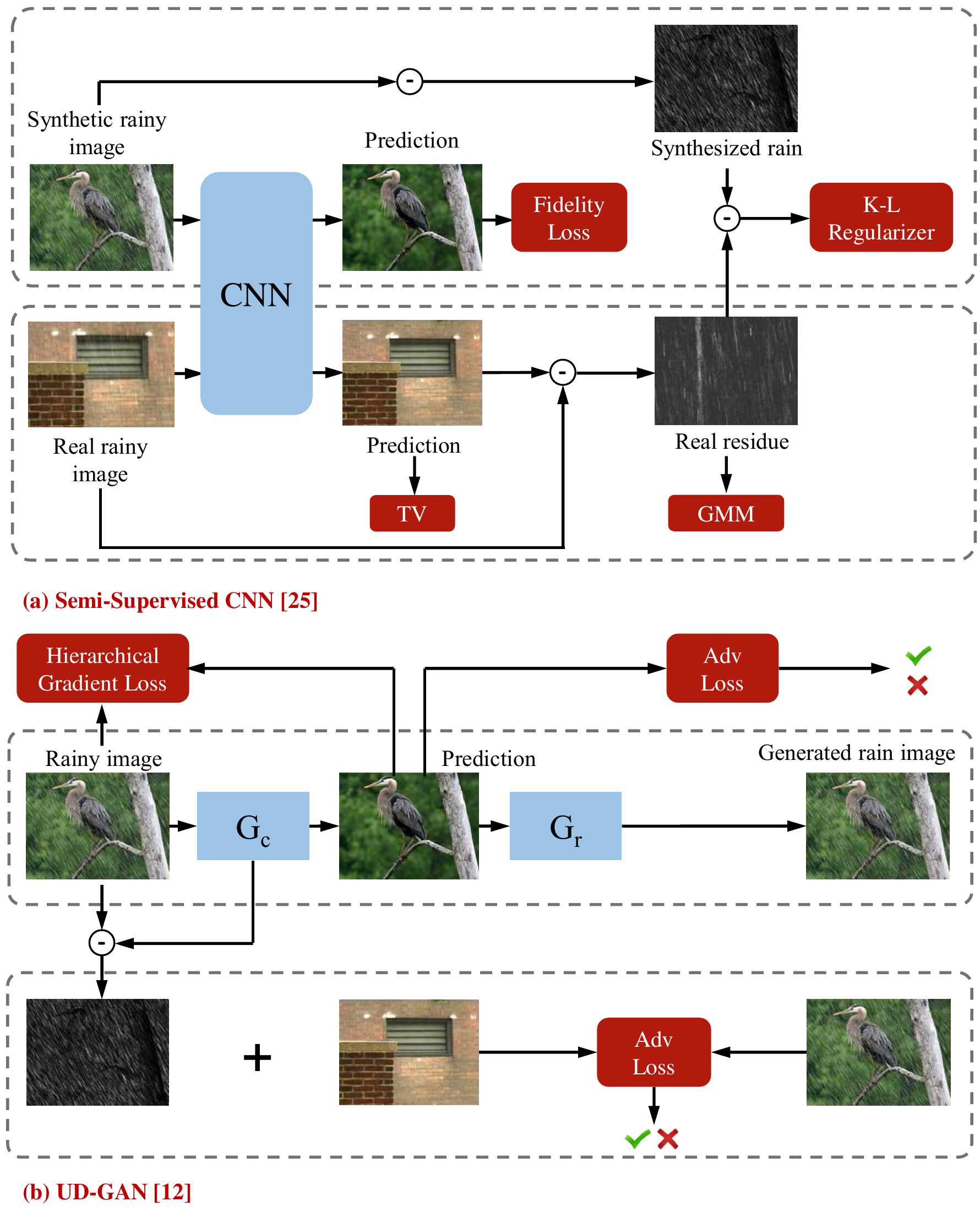}}
	\vspace{-4mm}
	\caption{Summary of semi/unsupervised learning-based rain removal methods.
		\wh{Semi/un-supervised learning methods make an attempt to improve the generality and scalability by learning  directly from real rain data.}}
	\vspace{-2mm}	
	\label{fig:unsupervised}
\end{figure}

\begin{figure*}[htbp]
	\centering
	\subfigure{
		\includegraphics[width=18cm]{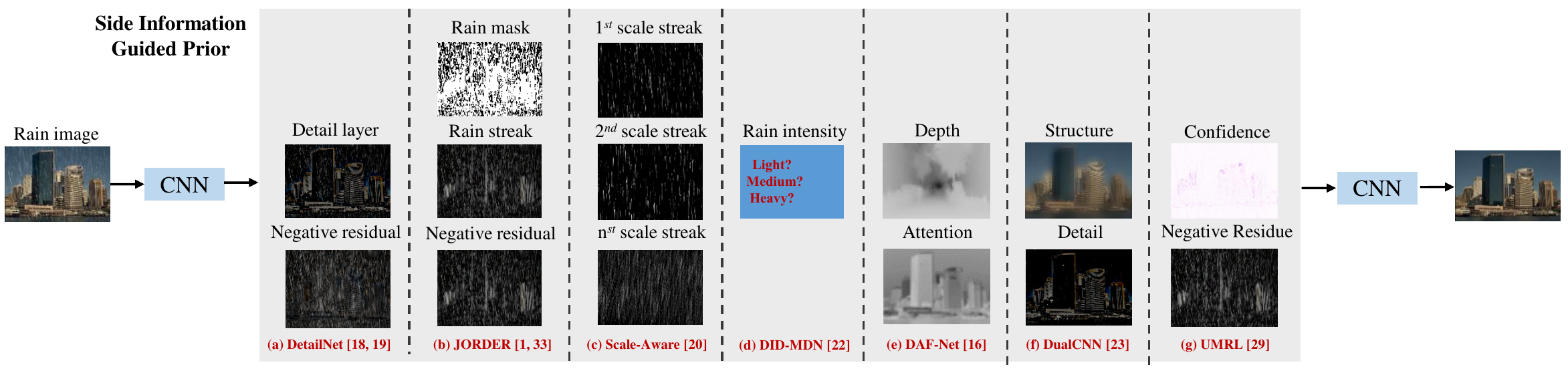}}
	\caption{Summary of the side information and priors for single-image rain removal.
		\wh{They injected into the networks to learn more effective features for deraining.}}
	\label{fig:side_info}
\end{figure*}

\vspace{0.4cm}

\noindent \textbf{Generative Adversarial Networks} 
To capture some visual properties of rain that
cannot be modeled and synthesized, the adversarial learning is introduced to 
reduce the domain gaps between the generated results  and real clean images.
The typical network architecture consists of two parts: generator and discriminator, where the discriminator attempts to assess whether a generated result is real or fake, which provides an additional feedback to regularize the generator to produce more visually pleasing results.
Zhang \textit{et al.}~\cite{CGAN_rain} directly apply the conditional generative adversarial network (CGAN) for the single image rain removal task, as shown in Fig.~\ref{fig:GAN}~(a).
CGAN is capable of capturing the visual properties beyond the signal fidelity, and presents results with better illumination, color and contrast distribution.
However, CGAN  sometimes might generate visual artifacts when the background of the testing rain image is different from those in the training set.

Li \textit{et al.}~\cite{Lil_2019_CVPR} propose a single-image deraining method that combines the physics-driven network and adversarial learning refinement network, as shown in Fig.~\ref{fig:GAN}~(c). 
The first stage learns from the synthesized data and estimates physics-related components, \textit{i.e.} rain streaks, the transmission, and the atmospheric light.
At the second refinement stage, a depth-guided GAN is proposed to compensate for the lost details and to suppress the introduced artifacts at the first stage.
Learning from real rain data, some visual properties of the results by these methods 
are significantly improved, namely removing rain accumulation more thoroughly and achieving a more balanced luminance distribution.
\wh{However, as GAN-based methods are not good at capturing fine-grained detail signals, 
the diversified appearances of real rain streaks  are also not properly modeled in these methods.}
\vspace{0.4cm}

\noindent \textbf{Semi/Unsupervised Learning Methods}
Recently, semi-supervised and unsupervised learning methods make an attempt to improve the generality and scalability by learning  directly from real rain data.
Wei \textit{et al}.~\cite{Wei_2019_CVPR} propose a semi-supervised learning method to make use of the priors in both synthesized paired data and unpaired real data, as shown in Fig.~\ref{fig:unsupervised}~(a).
In the proposed method, the residual is formulated as a specific parametrized rain streak distribution between an input rain image and its expected network output. The model trained on synthesized paired rain images is adapted to handle diversified rain in real scenarios with the guidance of the rain-streak distribution model. 
The method, however, does not show effective deraining results particularly for real rain images. This might be caused by the loose loss functions they impose to the network during the training process.

In~\cite{UD_GAN}, an unsupervised deraining generative adversarial network (UD-GAN) is proposed by introducing  self-supervised constraints, and  the intrinsic
priors extracted from unpaired rain and clean images, as shown in Fig.~\ref{fig:unsupervised}~(b).
Two collaborative modules are designed: One module is utilized to detect the difference between real rain images (real background images) and generated rain images (generated background images); while the other is introduced to adjust the luminance of the generated results, making the results more visually pleasing.
The method is capable of removing real rain from rain images, yet inevitably losing some details, particularly when the rain streaks are dense.

\vspace{0.4cm}

\noindent \textbf{Benchmark}
Li \textit{et al.}~\cite{Li_2019_CVPR} provided extensive study and evaluation of existing single image deraining algorithms with a newly proposed large-scale dataset including both synthetic and real-world rain images with  various rain types, \textit{i.e.} rain streak, raindrops, and mist. The benchmark also includes a wide range of evaluation criteria including the results of different methods quantitatively and qualitatively.


\subsubsection{Adherent Raindrop Removal}

Raindrops adhered to the camera lens can severely degrade visibility of a background scene in an image.
The goal of adherent raindrop removal is to detect and remove  raindrops from an input image.
Deraining is different from adherent raindrop removal, since rain images do not always suffer from adherent raindrop degradation, and vice versa: Adherent raindrop images do not always suffer from the degradation of rain streaks or rain accumulation. Nevertheless, we discuss it briefly here for the sake of the completeness of the survey. 

In~\cite{raindrop_stereo}, Yamashita \textit{et al.} develop a stereo system to detect and remove raindrops.
Subsequently, a method~\cite{raindrop_seq} is built based on the image sequence instead of stereo video. 
You \textit{et al.}~\cite{raindrop_you} propose a motion based 
method to detect raindrops, and apply video completion to remove the detected
regions.
Eigen~\textit{et al.}~\cite{Eigen_2013_ICCV} make the first attempt to tackle the problem of single-image raindrop removal.
A three-layer CNN is trained with pairs of raindrop degraded images and the corresponding clean ones.
It can handle relatively sparse and small raindrops as well as dirt, however, it fails to produce
clean results for large and dense raindrops.

Recently, Qian \textit{et al.}~\cite{Qian_2018_CVPR} develop an attentive GAN~(AttGAN)~\cite{Qian_2018_CVPR} by injecting  visual attention into both the generative and discriminative networks, as shown in Fig.~\ref{fig:GAN}~(b).
The visual attention does not only guide the discriminative network to focus more on local consistency of the restored raindrop regions, but also make the generative network pay more attention to the contextual information surrounding the raindrop areas.

\section{Technical Development Review}
In this section, we summarize the developments of existing deep-learning methods from the perspective of the network architectures, basic blocks, loss functions, and datasets.
These aspects significantly influence the network's learning capacity and thus determine the networks' deraining performance.

\begin{figure*}[htbp]
	\centering
	\subfigure{
		\includegraphics[width=18cm]{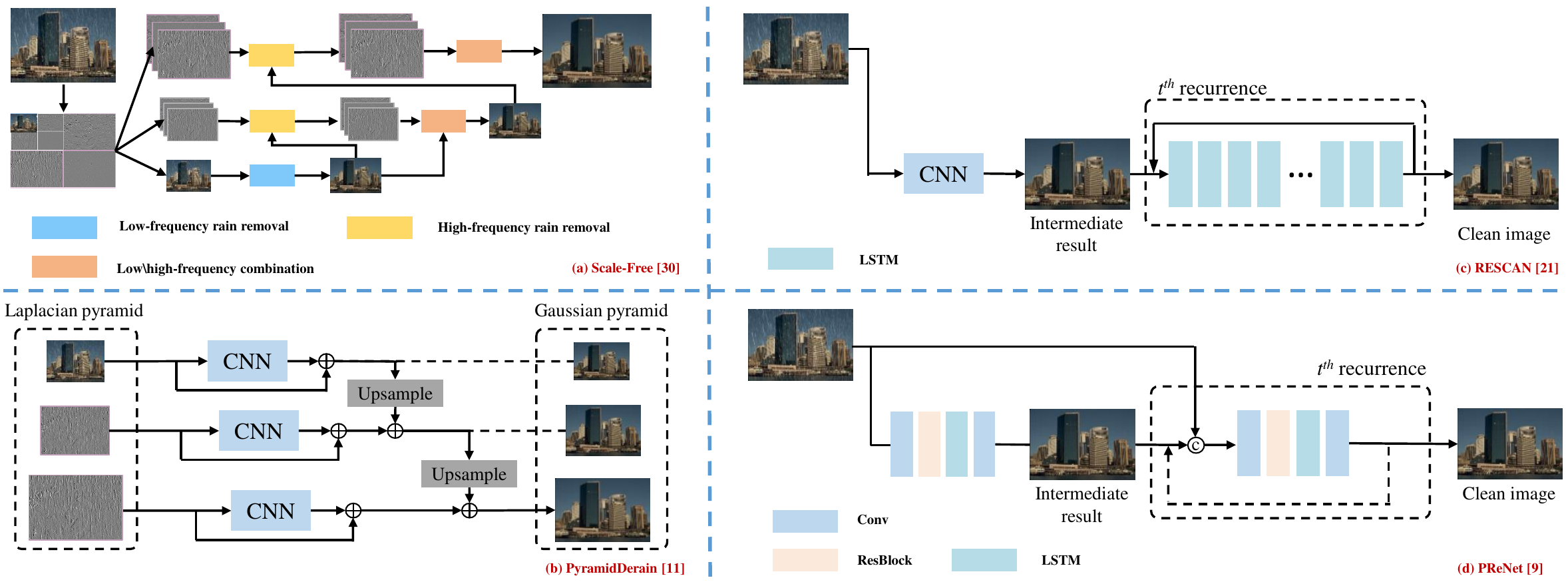}}
	\caption{Summary of the network improvement for single-image deraining.
	\wh{More effective network architectures are designed by relying on certain assumptions/constraints and general knowledge in image processing.}}
	\label{fig:network_architecture}
\end{figure*}

\subsection{Network Architectures}
\label{sec:network}

Since the publications of deep learning based deraining ~\cite{DetailNet2,2017_YangRainRemoval}, the successive methods aim to design more effective network architectures by relying on certain assumptions/constraints and general knowledge in image processing. In this section, we take a look at these developments.

\vspace{0.4cm}

\begin{table}[t]
	\centering
	\footnotesize
	\caption{Summary of side information used in previous works.}
	\label{tab:side}
	\begin{tabular}{cc}
		\hline
		Side information     & Methods                                           \\
		\hline
		Rain Mask            &  JORDER~\cite{2017_YangRainRemoval}               \\
		\hline
		Rain Density         &  JORDER~\cite{2017_YangRainRemoval}, DID-MDN~\cite{DID-MDI}            \\
		\hline
		Depth                &  DAF-Net~\cite{Hu_2019_CVPR}, HeavyRainRestorer~\cite{Lil_2019_CVPR}    \\
		\hline
		Attention            &  DAF-Net~\cite{Hu_2019_CVPR}, SPA-Net~\cite{Wang_2019_CVPR}, AttGAN~\cite{Qian_2018_CVPR}           \\
		\hline
		Intermediate Results &  JORDER-E~\cite{Yang_2019}, PReNet~\cite{Ren_2019_CVPR},   \\
		\hline
		Bands Results &   Scale-Free Rain Removal~\cite{Scale_free}, PyramidNet~\cite{PyramidNet}  \\
		\hline
	\end{tabular}
\end{table}

\noindent \textbf{Deraining Assumptions/Constraints}
Specific networks dedicated to certain problems usually perform better than generic networks. To create these specific networks, some constraints  or assumptions about the problems need to be injected. For deraining, these assumptions can relate to rain, background scenes, or other information.
By incorporating some of these assumptions, a network is expected to learn  the characteristics of rain better, and thus to  separate rain layer from the background layer more robustly.

Fu \textit{et al.}~\cite{DetailNet2} assume a rain image can be decomposed into detail and base layers, where the detail layer contains the image textures and rain streaks (Fig.~\ref{fig:side_info}~(a)), and the base layer  mostly contains the background and the rain accumulation. The proposed network thus attempts to derain the the detail layer, before fusing the layer with the base layer to obtain the rain-free output.  

Yang \textit{et al.}~\cite{2017_YangRainRemoval} construct a joint rain detection and removal network as shown in Fig.~\ref{fig:side_info}~(b)   to detect rain locations, estimate rain densities and predict rain sequentially, 
which boosts the capacity of the network to process rain and non-rain regions differently. 
Li \textit{et al}.~\cite{Li_scale} focus on the scale diversity of rain streaks. A scale-aware network as shown in Fig.~\ref{fig:side_info}~(c) consisting of parallel subnetworks is built to make it aware of different scales of rain streaks, producing better deraining performance for real images.
Zhang \textit{et al.}~\cite{DID-MDI} propose a density-aware rain removal method (DID-MDN) as shown in Fig.~\ref{fig:side_info}~(d) to automatically detect the rain-density as the guidance information for the successive deraining.

Aside from assumptions related to rain, some assumptions ar related to the background scenes.
Hu \textit{et al.}~\cite{Hu_2019_CVPR} analyze the complex visual effects in real rain and formulate a rain imaging model related to the scene depth. An end-to-end deep neural network as shown in Fig.~\ref{fig:side_info}~(e) is developed to extract depth-attentional features and to regress a residual map for predicting the clean image.
In~\cite{Pan_2018_CVPR}, a dual CNN as shown in Fig.~\ref{fig:side_info}~(f) is presented, where two branches learn the estimation of two parts of 
the target signal: structures and details. 

Another type of constraints is the confidence information of the residual between the estimated background layer and the ground-truth. In Fig.~\ref{fig:side_info}~(g), the model~\cite{Yasarla_2019_CVPR} 
learns about rain streaks by being guided by a per-pixel confidence map. This map is used to weigh wrongly estimated pixels in the back-propagation process, so that the network can pay more attention to these pixels during the training process. \wh{A summary of deraining side information used in previous works in provided in Table~\ref{tab:side}.}

\vspace{0.4cm}

\noindent \textbf{Image Processing Knowledge}
Some ideas present in the image processing literature can also useful in designing deraining network architectures, for instance: multi-scale structure, Laplacian pyramid, wavelet transform, \textit{etc.}.
A scale-free network~\cite{Scale_free} as shown in Fig.~\ref{fig:network_architecture}(a) pays attention to the scale variety of rain streaks in real scenes, and constructs a scale-free deraining architecture by unrolling a wavelet transform into a recurrent neural network, which can handle various kinds of rain at different scales. Guided by the hierarchical representation of the wavelet transform, a recurrent network consisting of two stages is built: 1) rain removal on the low-frequency component; 2) recurrent detail recovery on high-frequency components gudied by the recovered low-frequency component. 

PyramidDerain~\cite{PyramidNet} pursues a light-weighted pyramid of network as shown in Fig.~\ref{fig:network_architecture}(b) to remove rain from a single image. The decomposed Gaussian Laplacian image pyramid is combined with a deep network. The learning paradigm at each pyramid layer can be simplified, and the obtained network becomes shallow and has  less parameters. The model is quite light-weighed and achieves comparable state-of-the-art performance.

Li \textit{et al.}~\cite{RESCAN} propose a recurrent network to remove rain streaks progressively as shown in Fig.~\ref{fig:network_architecture}(c). 
The intermediate result from the last recurrence is taken as the input of the next recurrence, and the features are also forwarded and fused by RNN units, \textit{e.g.} GRU and LSTM, across recurrences.
Ren \textit{et al.}~\cite{Ren_2019_CVPR} utilize recursive computation to obtain more effective processing as shown in Fig.~\ref{fig:network_architecture}(d). The PReNet performs stage-wise operations that process the input and intermediate results to generate the clean output images progressively.

\begin{figure*}[htbp]
	\centering
	\subfigure{
		\includegraphics[width=18cm]{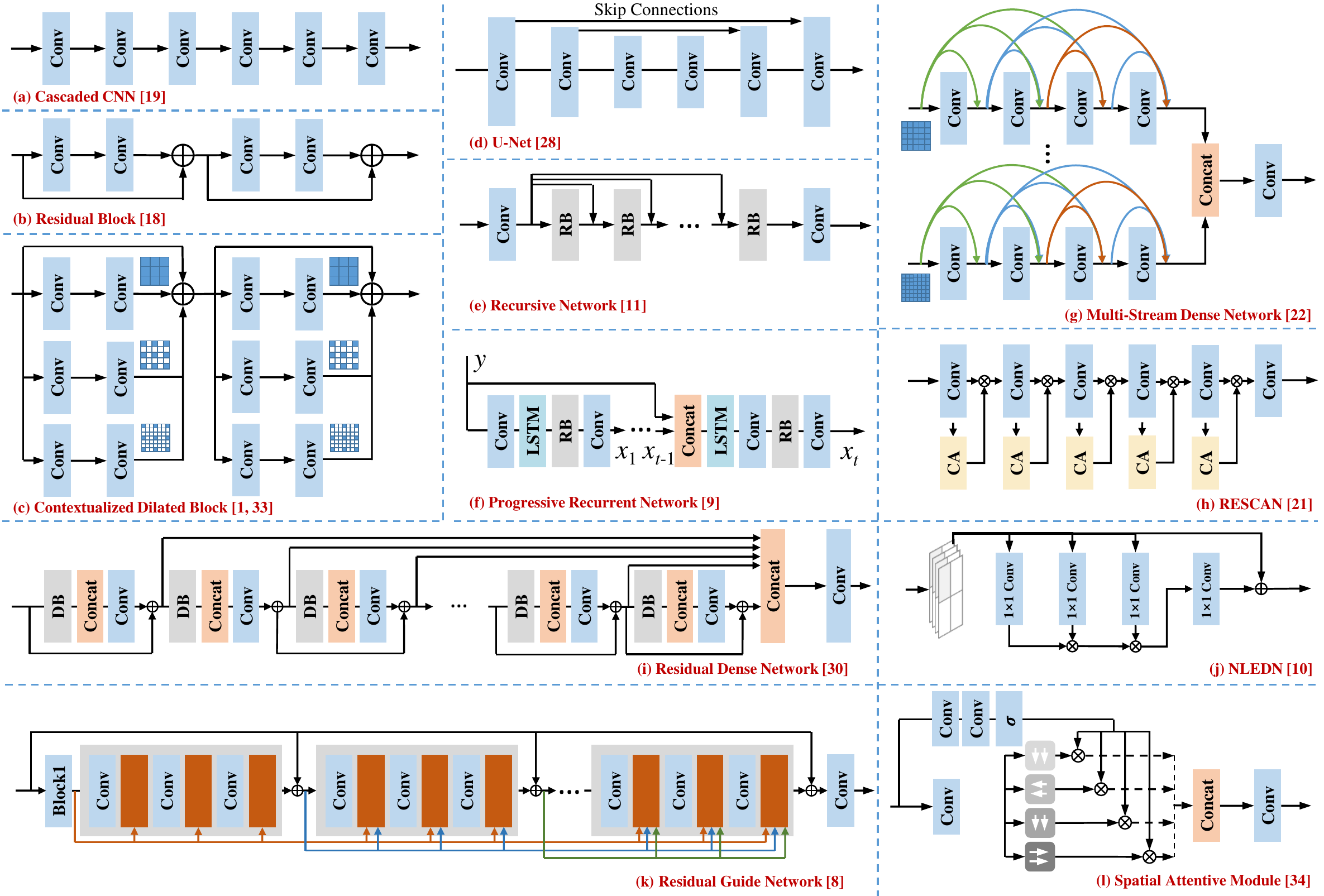}}
	\caption{The basic block improvement for single-image rain removal.
	\wh{The trend of newly proposed methods is to have more complex basic blocks with more powerful modeling capacities, which are further stacked into a more complex deraining network.}}
	\label{fig:basic_block}
\end{figure*}

\subsection{Basic Blocks}
With the development of deep-learning based methods, the trend of newly proposed methods is to have more complex basic blocks with more powerful modeling capacities, which are further stacked into a more complex deraining network.

\vspace{0.4cm}

\noindent \textbf{Using Existing Architectures}  DetailNet~\cite{DetailNet,DetailNet2} (Fig.~\ref{fig:basic_block} (a) and (b)) introduces a residual network and a cascaded CNN for rain removal. 
AttGAN~\cite{Qian_2018_CVPR} (Fig.~\ref{fig:basic_block} (d)) utilizes U-Net as the baseline of the generator, which is effective to fuse the information from different scales to obtain global information while maintaining local details.
RESCAN~\cite{RESCAN} (Fig.~\ref{fig:basic_block} (h)) introduces channel-wise attention to adjust the relative weighting among channels for better separating rain streaks from background layers.

Multi-stream dense network~\cite{DID-MDI} (Fig.~\ref{fig:basic_block} (g)) combines dense block and convolutional networks.
Residual dense network~\cite{Scale_free} (Fig.~\ref{fig:basic_block} (i)) integrates dense blocks into residual networks.
In~\cite{Residual_Guide} (Fig.~\ref{fig:basic_block} (k) and (e)), basic blocks are connected in the recursive way, where the input feature is also forwarded to the intermediate features of the network.
In~\cite{Ren_2019_CVPR} (Fig.~\ref{fig:basic_block} (f)), residual blocks are also aggregated in the recursive way and the LSTMs are  selected to connect different recurrences.

\vspace{0.4cm}

\noindent \textbf{Multi-Path Architectures}
One of the common architectures is the multi-path network. 
As shown in Fig.~\ref{fig:basic_block} (c), (g) and (l), the networks have different paths possessing different properties, \textit{i. e.} kernel sizes, dilation factors, and filter directions, to gather different kinds of information.
In Fig.~\ref{fig:basic_block} (c) and (e), different paths have different receptive fields, and thus can obtain both global information and maintain local structural details.
In Fig.~\ref{fig:basic_block} (l), the spatial redundancies are aggregated from different directions to form visual attention.

\vspace{0.4cm}

\noindent \textbf{Recursive Architectures}
In~\cite{Residual_Guide, RESCAN, Ren_2019_CVPR, PyramidNet}, recursive blocks are nested and  aggregated in the recursive way as shown in Fig.~\ref{fig:basic_block} (f), (k), (e) and (i).
The networks perform stage-wise operations that process the input and intermediate results to generate the output clean images progressively. 
Inter-stage recursive computation of different blocks is sometimes adopted to propagate information across the blocks.

Non-locally enhanced encoder-decoder network~\cite{NLEDN} as shown in Fig.~\ref{fig:basic_block} (j) incorporates nonlocal operations to the design of an end-to-end network for deraining. The non-local operation calculates the feature at a spatial position as a weighted sum of the features at a specific range of positions. In~\cite{Wang_2019_CVPR}, a spatial attentive module as shown in Fig.~\ref{fig:basic_block} (l) employs recurrent neural networks with ReLU and identity matrix initialization, to accumulate global contextual information in four directions. It utilizes another branch to capture the spatial contexts to selectively highlight the transformed rain features. 

\subsection{Loss Functions}

In existing deraining methods, several loss functions have been proposed to regularize the training of the deraining network.

\vspace{0.4cm}

\noindent \textbf{Fidelity-Driven Metrics}
Most studies need to use signal fidelity-driven matrices as the loss functions, such as Mean Squared Error~(MSE)~(L2), mean squared error~(MAE)~(L1), and SSIM~\cite{ssim}. They are defined as follows:
\begin{align}
	\label{eq:mse}
	\mathcal{L}_{\text{MSE}} \left(x, \hat{x}\right) & = \|  x - \hat{x} \|_2^2, \\
	\mathcal{L}_{\text{MAE}} \left(x, \hat{x}\right) & = \|  x - \hat{x} \|, \\
	\mathcal{L}_{\text{SSIM}} \left(x, \hat{x}\right) & =  \frac{ \left(2\mu_x \mu_{\hat{x}} + c_1 \right) \left(2\sigma_{x {\hat{x}}} + c_2 \right) }{  \left(\mu_x^2 + \mu_{\hat{x}}^2 + c_1 \right) \left(\sigma_{x}^2 + \sigma_{\hat{x}}^2 + c_2 \right)  },
\end{align}	
where $x$ and $\hat{x}$ are the ground truth and predicted clean images.
$\mu_x$ and $\mu_{\hat{x}}$ are the average of $x$ and $\hat{x}$, respectively.
$\sigma_x$ and $\sigma_{\hat{x}}$ are the variance of $x$ and $\hat{x}$, respectively.
$c_1$ and $c_2$ are two numbers to stabilize the division with weak denominator.

\vspace{0.4cm}

\noindent \textbf{Rain-Related Loss}
The rain-related variable prediction loss makes some outputs of the network 
predict the rain-related variable. 
For example, in~\cite{2017_YangRainRemoval}, the streak and the binary streak maps are connected to the corresponding losses as follows:
\begin{align}
	\mathcal{L}_{S} \left(s, \hat{s}\right) & = \|  s - \hat{s} \|_2^2, \\
	\mathcal{L}_{R} \left(r, \hat{r}\right) & = - \sum_{i} \left( \text{log} \hat{ {r}}_{i,1} {r}_{i,1} + \text{log} \left( 1 - \hat{ {r}}_{i, 2}\right)  \left( 1 -  {r}_{i,2} \right) \right), \nonumber
\end{align}
where $s$, $\hat{s}$, $r$ and $\hat{r}$ are 
the ground truth rain streak, predicted rain streak, ground truth rain mask, and predicted rain mask, respectively.
$i$ indexes the spatial pixel location.

\vspace{0.4cm}

\noindent \textbf{Multi-Scale Loss} 
The multi-scale loss~\cite{Qian_2018_CVPR} constrains the deraining network at different scales, which is expressed as:
\begin{equation}
	\mathcal{L}_{\text{MS}} \left(x, \hat{x}\right) = \sum_{i} \|  x_{s_i} - \hat{x}_{s_i} \|_2^2, 
\end{equation}
where $s_i$ indexes the scale, and $x_{s_i}$ and $\hat{x}_{s_i}$ are the down-sampled versions of $x$ and $\hat{x}$ with the scaling factor $s_i$.

\vspace{0.4cm}

\noindent \textbf{Perception-Driven Loss}
Applying perceptual and adversarial losses~\cite{Qian_2018_CVPR} improves the perceptual quality of generated results.
The perceptual loss is formulated as:
\begin{equation}
	\mathcal{L}_{\text{P}} \left(x, \hat{x}\right) =  \|  F\left( x \right) - F \left( \hat{x} \right) \|_2^2, 
\end{equation}
where $F(\cdot)$ is a pretrained CNN transformation.
The adversarial loss used for deraining network is represented as:
\begin{equation}
	\mathcal{L}_{\text{AdvGen}} = -\text{log} D \left( \hat{x} \right),
\end{equation}
where $D\left( \cdot \right)$ is a discriminator network that differentiates the generated $\hat{x}$ and the ground truth $x$.
A summary of loss functions used in previous works is given in Table~\ref{tab:loss}.

\begin{table}[t]
	\centering
	\footnotesize
	\caption{Summary of loss functions used in existing works.}
	\label{tab:loss}
	\begin{tabular}{cc}
		\hline
		Loss Function        &    Methods   \\
		\hline
		\hline
		MSE (L2)             &   \tabincell{c}{JORDER~\cite{2017_YangRainRemoval,Yang_2019}    DetailNet~\cite{DetailNet,DetailNet2}, DID-MDN~\cite{DID-MDI}, \\ DAF-Net~\cite{Hu_2019_CVPR}, CGAN~\cite{CGAN_rain}, DualCNN~\cite{Pan_2018_CVPR}, \\ PReNet~\cite{Ren_2019_CVPR}, RESCAN~\cite{RESCAN}, Scale-Free~\cite{scale_aware}, \\ Residual guided net~\cite{Residual_Guide}, Semi-Supervised~\cite{UD_GAN}, \\ Scale-Aware~\cite{scale_aware}}  \\
		\hline
		MAE (L1)             &    \tabincell{c}{PyramidDerain~\cite{PyramidNet}, NLEDN~\cite{NLEDN}, SPA-Net~\cite{Wang_2019_CVPR}, \\ UD-GAN~\cite{UD_GAN}}                \\
		\hline
		SSIM                 &    \tabincell{c}{PyramidDerain~\cite{PyramidNet}, PReNet~\cite{Ren_2019_CVPR}, SPA-Net~\cite{Wang_2019_CVPR}, \\  Residual guided net~\cite{Residual_Guide} }              \\
		\hline
		Adversarial          &    AttGAN~\cite{Qian_2018_CVPR}, CGAN~\cite{CGAN_rain}                   \\
		\hline
		Perceptual           &    AttGAN~\cite{Qian_2018_CVPR}, DID-MDN~\cite{DID-MDI}, CGAN~\cite{CGAN_rain}    \\
		\hline
		Multi-Scale          &    AttGAN~\cite{Qian_2018_CVPR}                   \\
		\hline
		Variable             &  \tabincell{c}{JORDER~\cite{2017_YangRainRemoval,Yang_2019}, Semi-Supervised~\cite{UD_GAN},\\
		DID-MDN~\cite{DID-MDI},  DAF-Net~\cite{Hu_2019_CVPR}, AttGAN~\cite{Qian_2018_CVPR},\\ Scale-Aware~\cite{scale_aware}, SPA-Net~\cite{Wang_2019_CVPR}, UD-GAN~\cite{UD_GAN}}     \\
		\hline
	\end{tabular}
\end{table}

\begin{table*}[t]
	\centering
	\caption{Summary of datasets used in previous works.}
	\label{tab:summary_dataset}
	\begin{tabular}{c|clcc}
		\hline
		Dataset     & \tabincell{c}{Number \\ (\#train/\#test)}   & \qquad \qquad \qquad \qquad \qquad \qquad Highlight & Rain Model & Publication \\
		\hline
		\hline
		Rain12      & 12       &  \tabincell{l}{Only for testing.}  & ACM & Li \textit{et al.}~\cite{li2016rain}   \\
		\hline
		Rain100L    & 1,800/100 &  \tabincell{l}{Synthesized with only one type of rain streaks (light rain case).} & ACM & Yang \textit{et al.}~\cite{2017_YangRainRemoval}   \\
		\hline
		Rain100H    & 1,800/100 &  \tabincell{l}{Synthesized with five types  of  rain streaks (heavy rain case).} & ACM & Yang \textit{et al.}~\cite{2017_YangRainRemoval}   \\
		\hline
		Rain800     & 700/100  & \tabincell{l}{Clean images are selected from BSD500 and UCID~\cite{UCID}.} & ACM & Zhang \textit{et al.}~\cite{CGAN_rain}  \\
		\hline
		Rain14000   & 9,100/4,900  &  \tabincell{l}{1,000 clean image used to synthesize 14,000 rain images.}  & ACM & Fu \textit{et al.}~\cite{DetailNet}   \\
		\hline
		Rain12000   & 12,000/4000        &   \tabincell{l}{The data has three kinds of densities.}      & ACM & Zhang \textit{et al.}~\cite{DID-MDI}  \\ \hline
		RealDataset & 28,500/1,000 &  \tabincell{l}{The ground truth data is  synthesized based on temporal redundancy \\ and visual properties.}  & ACM & Wang \textit{et al.}~\cite{Wang_2019_CVPR}   \\ \hline
		NYU-Rain & 13,500/2,700 & \tabincell{l}{Background images and the depth information are selected from \\ NYU-Depth V2~\cite{Silberman_ECCV12}.} & HRM & Li \textit{et al.}~\cite{Lil_2019_CVPR} \\ \hline 
		Outdoor-Rain & 9,000/1,500 & \tabincell{l}{The background images are collected from~\cite{Qian_2018_CVPR}, and the depth \\ information is estimated by~\cite{monodepth17}.} & HRM &  Li \textit{et al.}~\cite{Lil_2019_CVPR} \\ \hline 
		
		RainCityscapes & 9,432/1,188  &  \tabincell{l}{The rain-free images are selected from the training and validation \\ sets of Cityscape~\cite{Cordts2016Cityscapes}. Rain patches are selected from~\cite{li2016rain}.}
		& DARM & Hu \textit{et al.}~\cite{Hu_2019_CVPR}   \\ \hline		
		MPID        &  1,561/419  &  \tabincell{l}{The MPID dataset covers a much larger diversity of rain models,\\  including both synthetic and real-world   images, serving both human \\ and machine visions.}
		& ACM + HRM & Li \textit{et al.}~\cite{Li_2019_CVPR}   \\

		\hline
	\end{tabular}
\end{table*}

\subsection{Datasets}
There are a few benchmarking datasets for image deraining, as introduced in Table~\ref{tab:summary_dataset}. These datasets are useful for network training as well as for evaluation:
\begin{itemize}
\item \textbf{Rain12}~\cite{li2016rain} includes
12 synthesized rain images with only one type of rain streaks.

\item 
\textbf{Rain100L} and \textbf{Rain100H}~\cite{2017_YangRainRemoval} include the synthesized rain images with only one type and five types of rain streaks, respectively.

\item 
\textbf{Rain800}~\cite{CGAN_rain}'s
training set consists of a total of 700 images, where 500 images are randomly chosen
from the first 800 images in the UCID dataset~\cite{UCID} and 200 images are randomly chosen from the BSD500's training set~\cite{amfm_pami2011}. 
The testing set consists of a total of 100 images, where 50 images are randomly chosen from the last 500 images
in the UCID dataset and 50 images are randomly chosen from the testing set of the BSD-500 dataset.

\item 
\textbf{Rain14000}~\cite{DetailNet}  includes 1000 clean images from UCID dataset~\cite{UCID},
BSD dataset~\cite{amfm_pami2011} and Google image search being used to synthesize
rainy images.

\item 
\textbf{Rain12000}~\cite{DID-MDI} consists of 12,000 images in the training set, where each image
is assigned a label based on its corresponding rain-density
level (\textit{i.e.} light, medium and heavy). 
There are 4,000 images per rain-density level in the dataset. 
The synthesized testing set includes 1,200 images.

\item 
\textbf{RealDataset}~\cite{Wang_2019_CVPR} includes 29,500 rain/rain-free image pairs that cover a wide range of natural rain scenes, where the rain-free images are synthesized based on temporal redundancy and visual properties.

\item 
NYU-Rain~\cite{Lil_2019_CVPR} is a new synthetic rain dataset taking images from NYU-Depth V2~\cite{Silberman_ECCV12} as background and the provided depth information to generate rain streak and accumulation layers. The dataset also considers the effect of image blurring presented in the rain image. It contains 16,200 image samples, where 13,500 images are used as the training set.

\item 
\textbf{Outdoor-Rain}~\cite{Lil_2019_CVPR}'s background images are collected from~\cite{Qian_2018_CVPR}, and the depth information used to synthesize accumulation is produced by~\cite{monodepth17}. The dataset includes 9000 training images and 1,500 validation images.

\item 
\textbf{MPID}~\cite{Li_2019_CVPR}'s training set includes 
2400 synthetic rain streak image pairs, 861 synthetic raindrop image pairs, and 700 synthetic rain and mist image pairs.
The testing set includes 200 synthetic rain streak image pairs, 149 synthetic raindrop image pairs, and 70 synthetic rain and mist image pairs, as well as 50 real rain streak images, 58 real raindrop images, and 30 real rain and mist images.
The testing set also includes 2,496 and 2,048 real captured images in the driving and surveillance video conditions with human annotated object bounding boxes.

\item 
\textbf{RainCityscapes}~\cite{Hu_2019_CVPR} consists of  
262 training images and 33 testing images from the training and validation
sets of Cityscape~\cite{Cordts2016Cityscapes}, which  are selected as the clean background images. 
Rain patches are selected from~\cite{li2016rain}. 
There are in total 9,432 training images and 1,188 testing images.
\end{itemize}

\begin{figure*}[t]
	\centering
	\subfigure{
		\includegraphics[width=5.9cm]{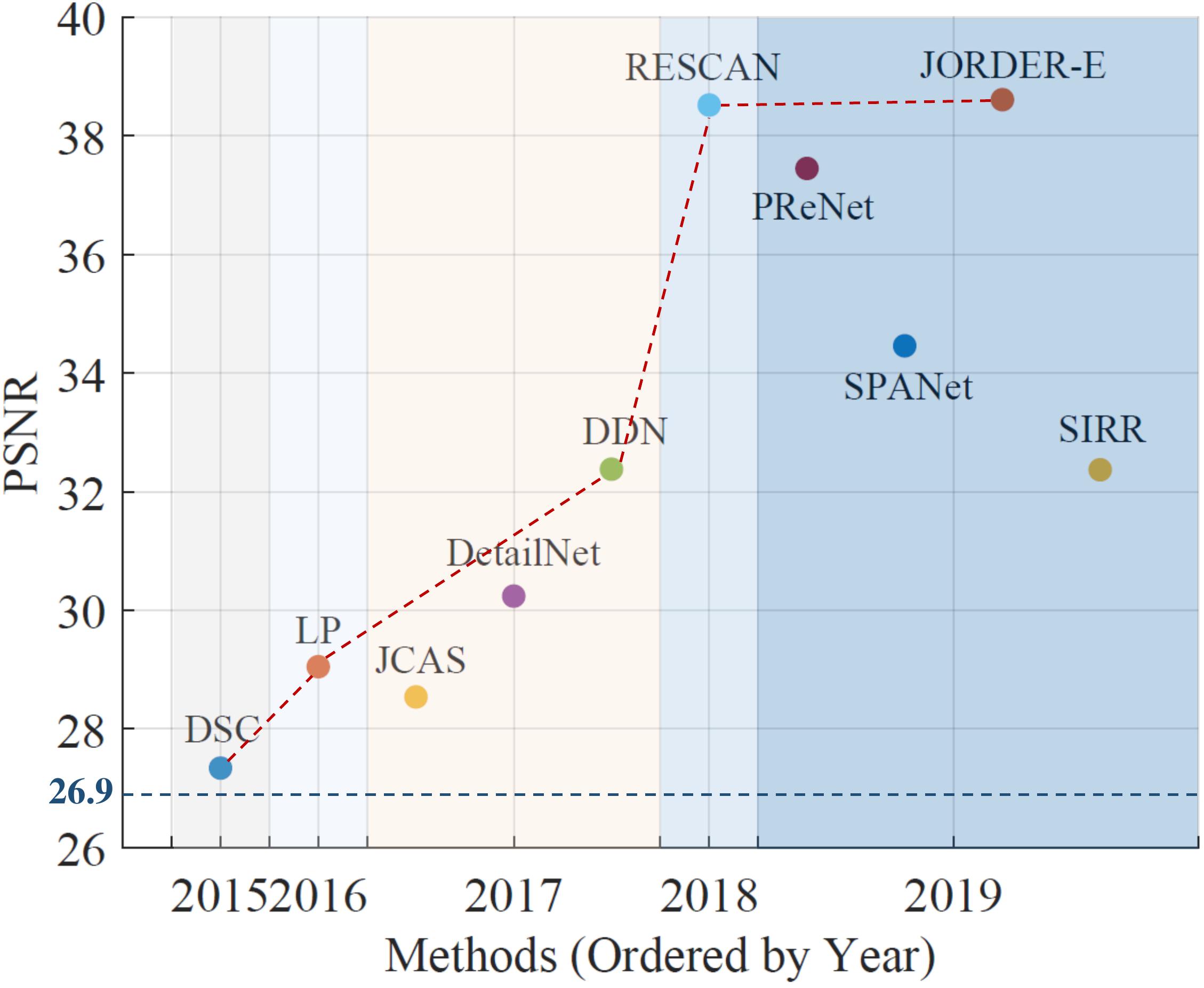}} 
	\hspace{-1.5mm}
	\subfigure{
		\includegraphics[width=5.9cm]{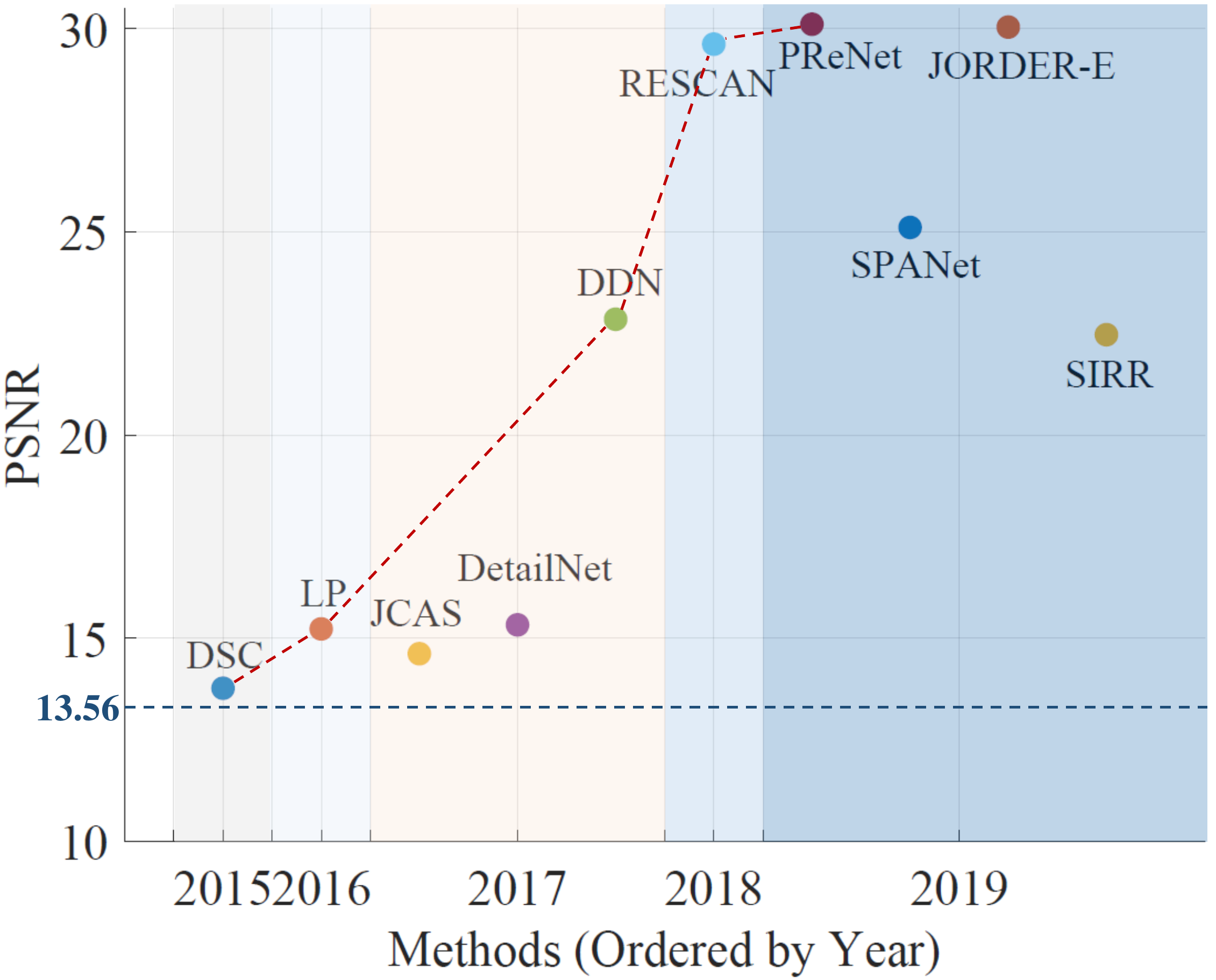}}
	\hspace{-1.5mm}	
	\subfigure{
		\includegraphics[width=5.9cm]{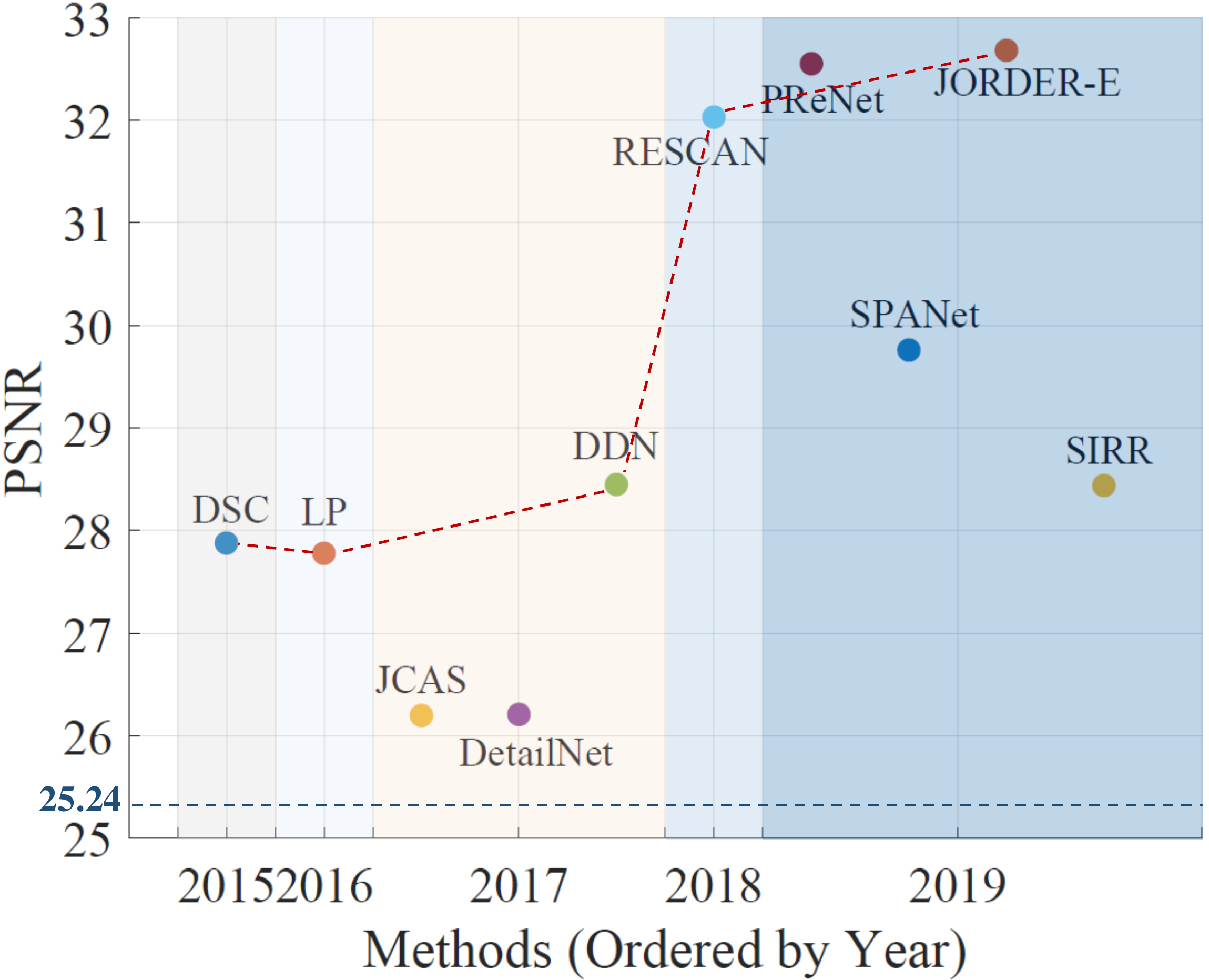}}
	\\
	\setcounter{subfigure}{0}
	\subfigure[Rain100L]{
		\includegraphics[width=5.9cm]{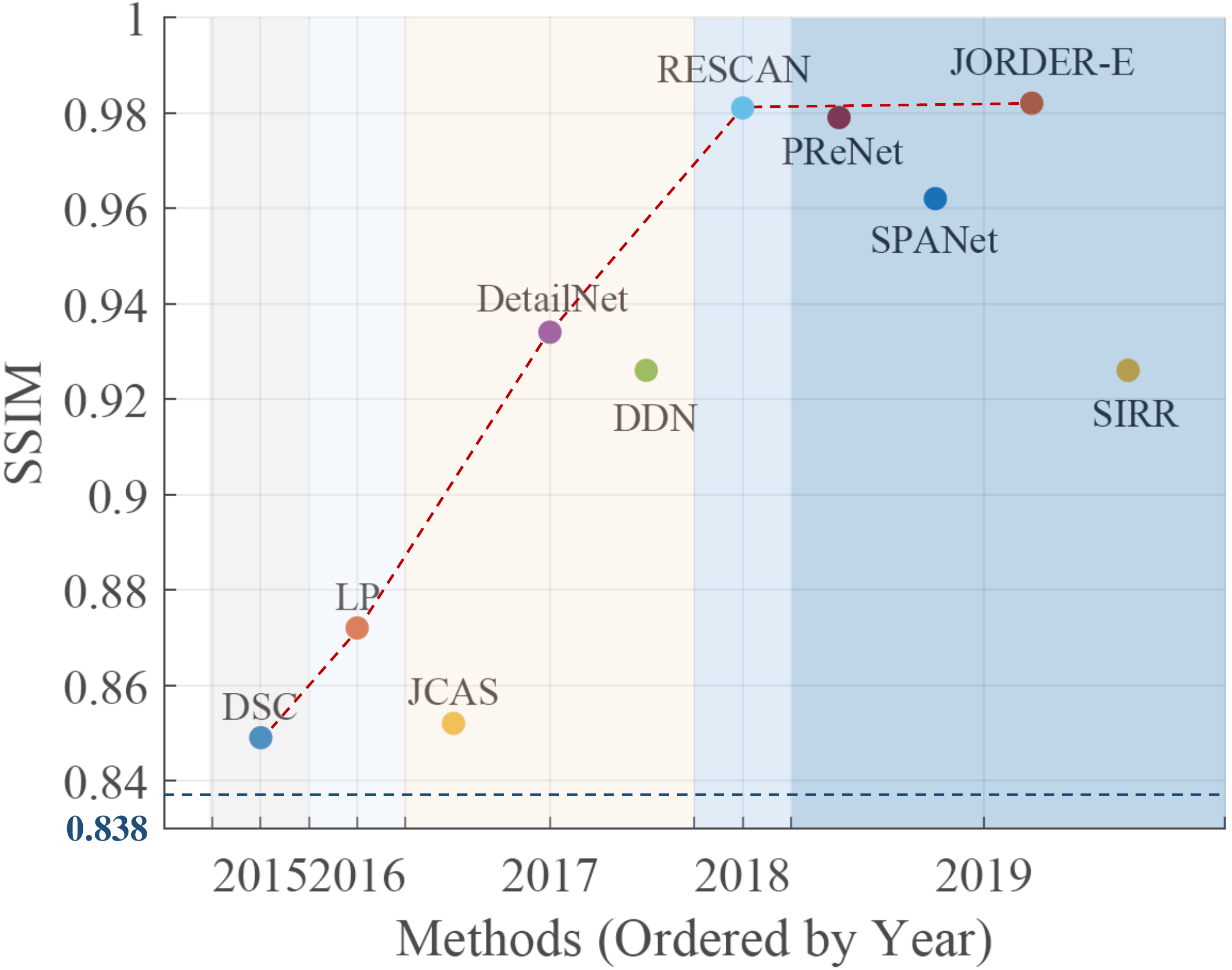}}
	\subfigure[Rain100H]{
		\includegraphics[width=5.9cm]{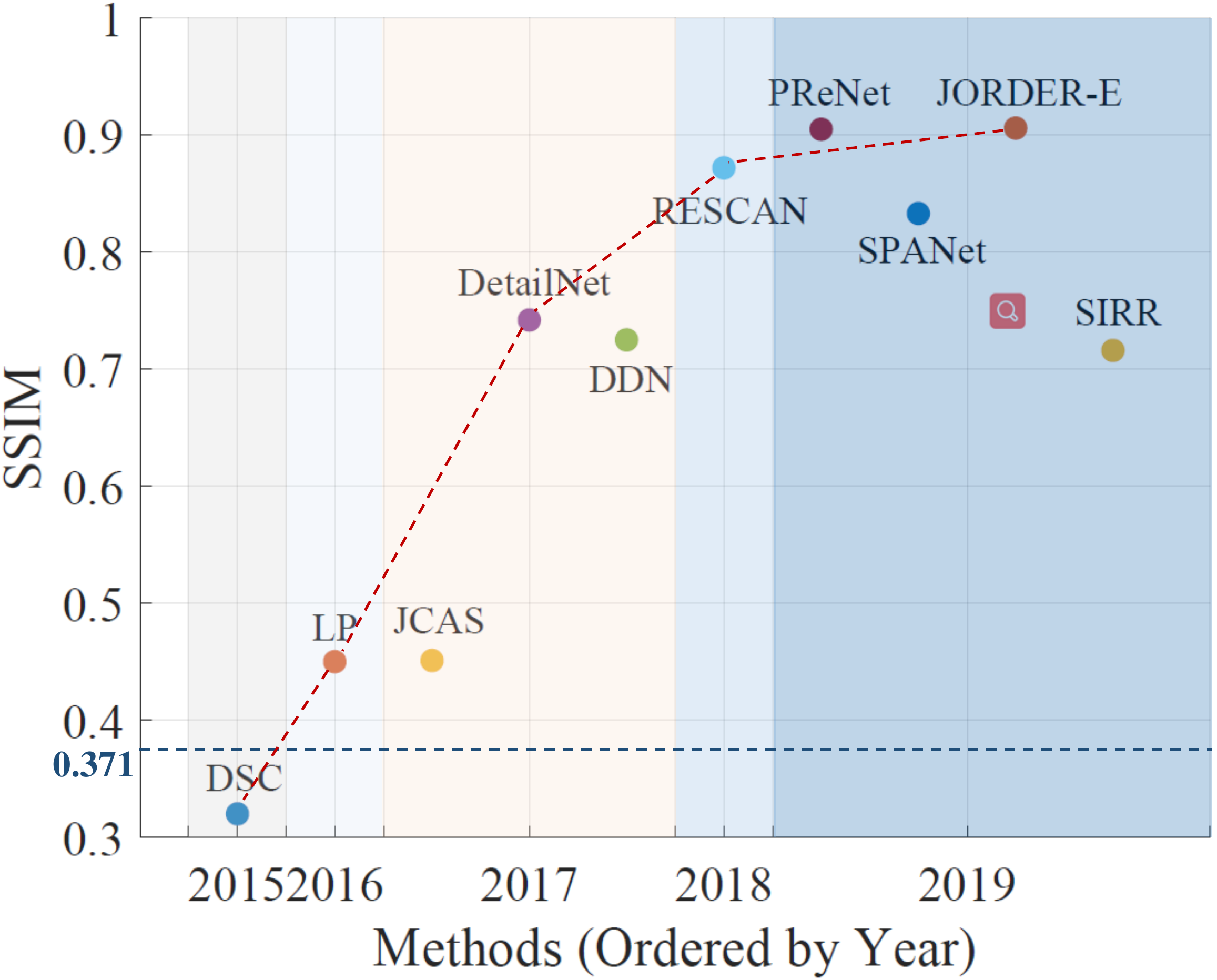}}
	\subfigure[Rain1400]{
		\includegraphics[width=5.9cm]{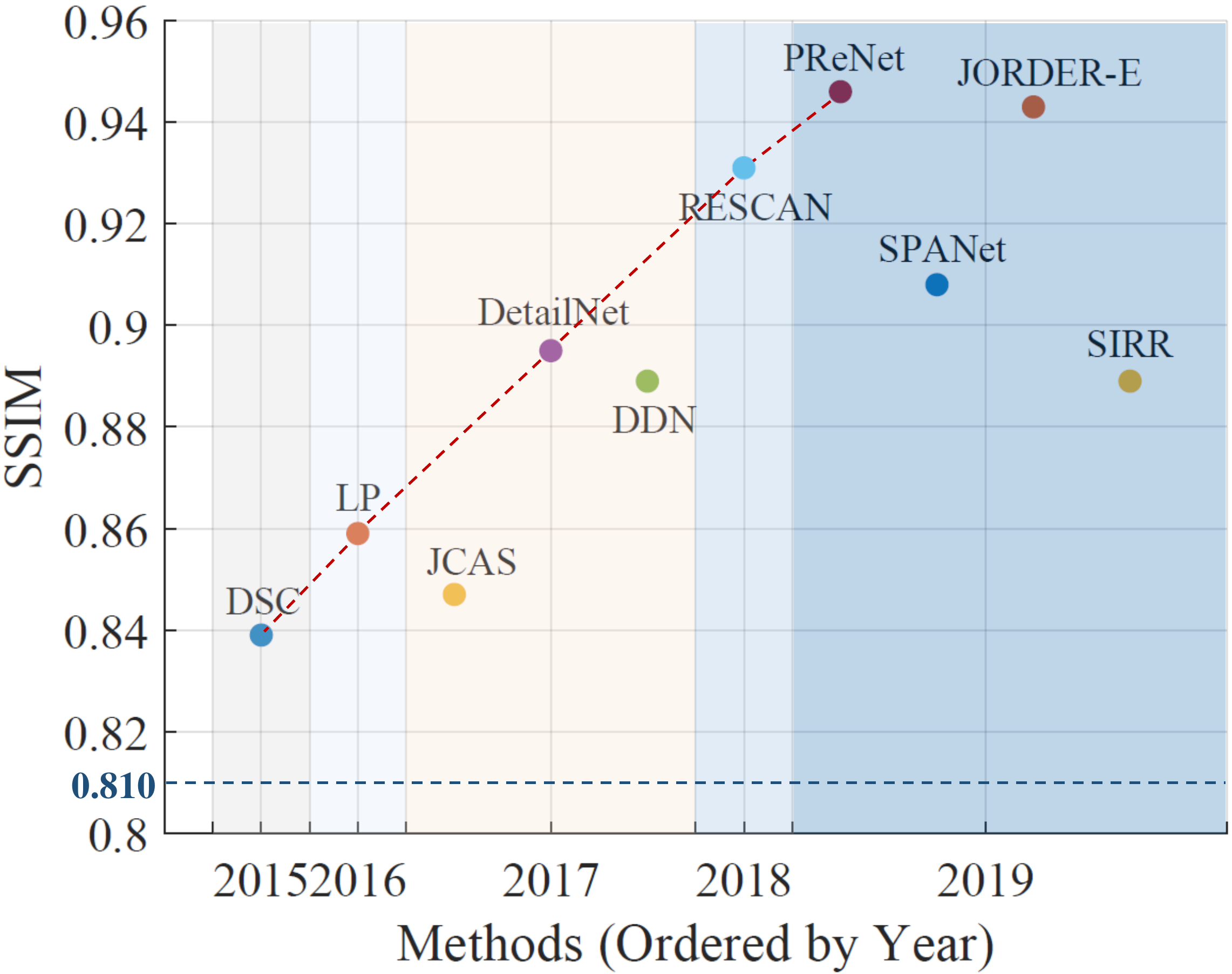}}
	\caption{The objective results of different methods. Top panel: PSNR. Bottom panel: SSIM. 
	All methods are sorted by year. A red curve connects the top performance from 2015 to 2019. It is shown that, the objective performance gains converge gradually. 
	}
	\vspace{-2mm}
	\label{fig:rain_results}
\end{figure*}

\begin{table*}[htbp]
	\centering
	\scriptsize
	\caption{The non-reference metric results of different methods. Heavy denotes HeavyRainRemoval. 
	{\color{red}Red}, {\color{blue}blue}, and {\color{green}green} denote the best, second best, and third best results.
	}
	\label{tab:nr_result}
	\vspace{-3mm}
	\begin{tabular}{cccccccccccccc}
		\hline
		Methods   & Input  & ID     & DSC    & LP     & DetailNet & Heavy & DID-MDN & RESCAN & JCAS   & JORDER-E & PReNet & SPANet & $\uparrow$ or $\downarrow$  \\
		\hline
		\hline
		NIQE      & 5.38   & 4.46   & 4.55   & 5.46   & 5.34      & 4.97      & 5.07  & {\color{red}3.78}   & 4.97   & {\color{green}3.93}     & 4.78   & {\color{blue}3.92}   & $\downarrow$ \\
		PIQE      & 38.36  & 39.41  & 40.04  & 64.86  & 36.50     & 52.81     & 38.97 & {\color{red}24.28}  & 35.64  & {\color{blue}29.77}    & 45.72  & {\color{green}30.40}  & $\downarrow$ \\
		BRISQUE   & 34.00  & 31.44  & 32.51  & 41.01  & 32.99     & 37.32     & 30.12 & {\color{red}25.70}  & 32.91  & {\color{blue}26.51}    & 30.28  & {\color{green}28.38}  & $\downarrow$ \\
		ILNIQE    & 31.09  & 31.20  & 29.16  & 41.78  & 30.72     & 31.84     & {\color{red}26.32} & 30.63  & {\color{blue}26.53}  & {\color{green}28.46}    & 29.70  & 30.58  & $\downarrow$ \\
		SSEQ      & 24.91  & 28.66  & 29.02  & 44.84  & {\color{green}22.83}     & 32.76     & 26.41 & {\color{blue}21.94}  & {\color{red}17.94}  & 23.27    & 26.48  & 25.15  & $\downarrow$ \\
		SR-Metric & 7.90   & 7.73   & 7.46   & 4.76   & {\color{blue}8.12}      & 7.26      & {\color{green}8.06}  & 7.90   & {\color{red}8.29}   & 7.82     & 7.66   & 7.85   & $\uparrow$  \\
		ENIQA     & 0.1508 & 0.2012 & 0.1886 & 0.2631 & {\color{blue}0.1323}    & 0.2091    & {\color{green}0.1331} & 0.1347 & {\color{red}0.1166} & 0.1394   & 0.1499 & 0.1445 & $\downarrow$ \\
		BIQAA     & 0.0107 & {\color{blue}0.0036} & 0.0051 & {\color{red}0.0031} & 0.0123    & 0.0069    & 0.0085 & 0.0041 & 0.0165 & {\color{green}0.0040}   & 0.0043 & 0.0078 & $\downarrow$  \\
		BIQI      & {\color{red}42.84}  & 11.75  & 34.12  & 18.58  & {\color{blue}40.05}     &-5.15    & {\color{green}37.05} & 35.40  & 29.81  & 34.46    & 31.02  & 24.34  & $\uparrow$ \\
		BLIINDS-II  & 22.55  & {\color{green}19.80}  & 22.00  & 22.85  & 21.78     & 25.33     & 20.60 & 12.58  & 20.13  & {\color{red}12.53}    & 24.03  & {\color{blue}16.08}  & $\downarrow$  \\
		FRISQUE   & 54.99  & 54.32  & 52.95  & 38.40  & {\color{blue}58.97}     & 52.75     & 25.29 & 57.56  & {\color{red}66.70}  & 58.49    & {\color{green}58.95}  & 58.54  & $\uparrow$  \\
		\hline
        Average MOS       & - & 	0.1042 &	0.2232 &	0.2440 &	{\color{green}1.8377} &	1.0974 &	0.4004 &	1.2413 &	0.3527 &	{\color{red}3.1366} &	{\color{blue}2.3252} 	& 1.0 &  $\uparrow$ \\
		MOS & -  &  0.1466  &  0.2804  &  0.3050  &  {\color{green}1.8897}  &  1.2510 & 0.4825  &  1.3037  &  0.4233  &  {\color{red}3.1988}  &  {\color{blue}2.3802}  &  1.0  &  $\uparrow$ \\
		\hline
	\end{tabular}
\end{table*}

\section{Performance Summary}

We select a number of recent deraining algorithms from different categories to be evaluated:
\begin{enumerate}
	\item 
Image Decomposition, ID~\cite{ID},
	\item 
Discriminative Sparse Coding, DSC~\cite{luo2015removing},
	\item 
Gaussian mixture model Layer Prior, LP~\cite{li2016rain},
	\item 
Joint Convolutional Analysis and Synthesis Sparse  
Representation, JCAS~\cite{jcas},
	\item 
Deep Detail Network, DetailNet~\cite{DetailNet2},
DDN~\cite{DetailNet},
	\item 
Directional Global Sparse Model, DGSM~\cite{DENG2018662},
	\item 
Recurrent Squeeze-and-Excitation Context Aggregation Net, RESCAN~\cite{RESCAN},
	\item 
Progressive Recurrent Network, PReNet~\cite{Ren_2019_CVPR},
	\item 
Enhanced JOint Rain DEtection and Removal, JORDER-E~\cite{Yang_2019},
	\item 
Heavy Rain Image Restoration, HeavyRainRestorer~\cite{Lil_2019_CVPR},
	\item 
Spatial Attentive Network, SPANet~\cite{Hu_2019_CVPR},
	\item 
Semi-supervised Image rain Removal, SSIR~\cite{Wei_2019_CVPR}, and 
	\item 
Density-aware Image De-raining using Multi-stream Dense Network, DID-MDN~\cite{DID-MDI}.
\end{enumerate}

LP is built based on Gaussian mixture models.
ID, DSC, and JCAS are designed based on sparse coding.
JORDER-E, DetailNet, DDN, DID-MDN, 
PReNet, SPANet, and RESCAN are deep-learning based methods.
HeavyRainRestorer integrates deep CNN and generative adversarial learning for rain removal.
In our experiments, JORDER-E, DetailNet, PReNet, and RESCAN are trained on Rain100H.
SPANet is trained on RealDataset.
DID-MDN is is trained on Rain800.
HeavyRainRestorer is trained on NYU-Rain and Outdoor-Rain.

Peak signal-to-noise ratio (PSNR) and SSIM~\cite{Brooks08structuralsimilarity} are used for performance evaluation.
A few different metrics are used, particularly when we do not have ground-truths, which we call non-reference metrics: 
Naturalness Image Quality Evaluator (NIQE)~\cite{niqe},
Perception-based Image Quality Evaluator (PIQE)~\cite{piqe},
Blind/Referenceless Image Spatial Quality Evaluator (BRISQUE)~\cite{BRISQUE},
Integrated Local NIQE (IL-NIQE)~\cite{ILNIQE},
Spatial-Spectral Entropy based Quality (SSEQ)~\cite{LIU2014856},
SR Metrics~\cite{SR_metrics},
Entropy-based No-reference Image Quality Assessment (ENIQA)~\cite{Chen2019},
Blind Image Quality Assessment through Anisotropy (BIQAA)~\cite{Gabarda_07},
and Blind Image Quality Assessment (BIQA)~\cite{BIQA},
BLind Image Integrity Notator using DCT Statistics (BLIINDS-II)~\cite{BLIINDS}.
These metrics measure the visual quality in different ways including  human perception in lightness distortion, texture preservation, spatial domain statistics, and natural preservation, \textit{etc}.

\subsection{Quantitative Evaluation}
We compare the quantitative results of different rain removal methods in Fig.~\ref{fig:rain_results}.
The numbers  are obtained from~\cite{2019arXiv190908326W}.
To observe the trend of the performance changes over the years, we order different methods by year in Fig.~\ref{fig:rain_results}.
This figure provides some interesting information.
First, most of the deep learning-based methods achieve significantly superior performance to model-based methods. For example, DDN obtains more than 3dB, 7dB, and 0.7dB on Rain100L, Rain100H, and Rain1400, respectively.
Second, the best performance of different methods gradually converges. The performance gaps between RESCAN, PReNet and JORDER-E are considerably close.

\subsection{Qualitative  Evaluation}
\label{sec:subjective}
\begin{figure*}[htbp]
	\flushright
	\subfigure{
		\includegraphics[width=2.9cm]{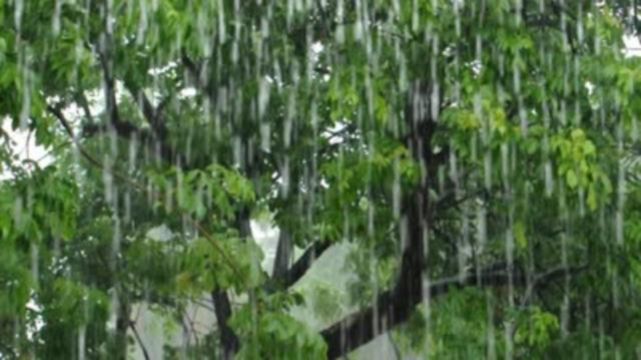}} 
	\hspace{-1.9mm}
	\subfigure{
		\includegraphics[width=2.9cm]{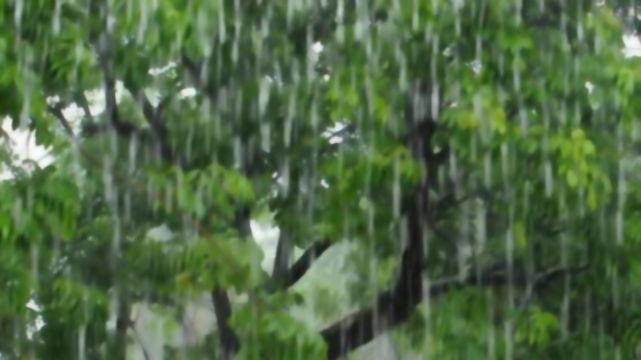}}
	\hspace{-1.9mm}	
	\subfigure{
		\includegraphics[width=2.9cm]{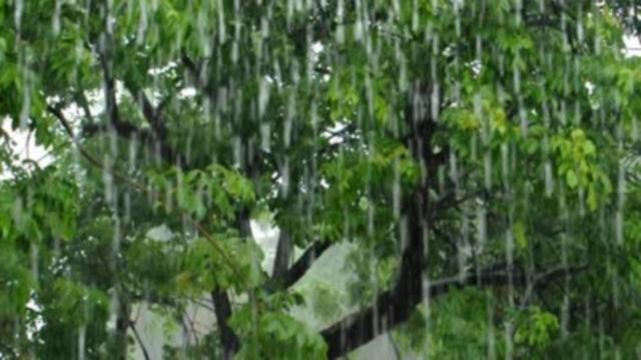}}
	\hspace{-1.9mm}	
	\subfigure{
		\includegraphics[width=2.9cm]{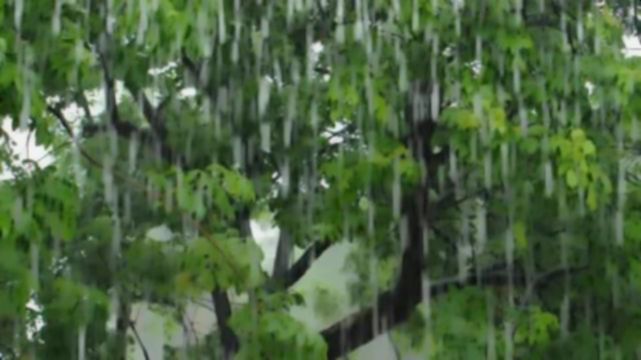}}
	\hspace{-1.9mm}	
	\subfigure{
		\includegraphics[width=2.9cm]{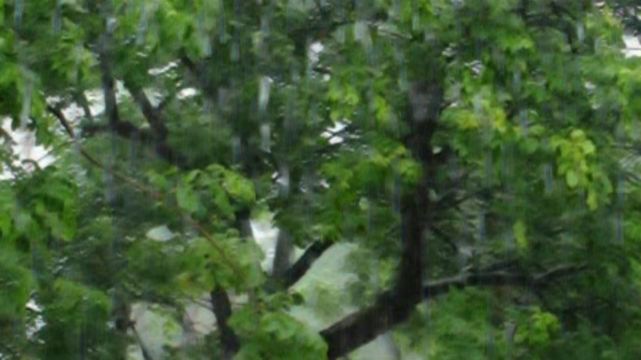}}
	\hspace{-1.9mm}	
	\subfigure{
		\includegraphics[width=2.9cm]{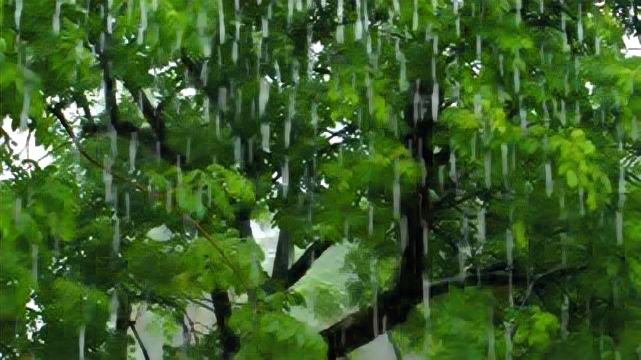}}
	\\ \vspace{-2mm}
	\subfigure{
		\includegraphics[width=2.9cm]{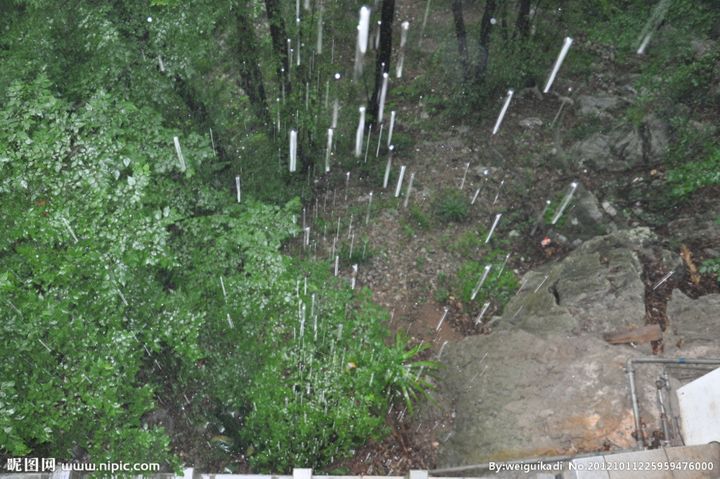}}
	\hspace{-1.9mm}
	\subfigure{
		\includegraphics[width=2.9cm]{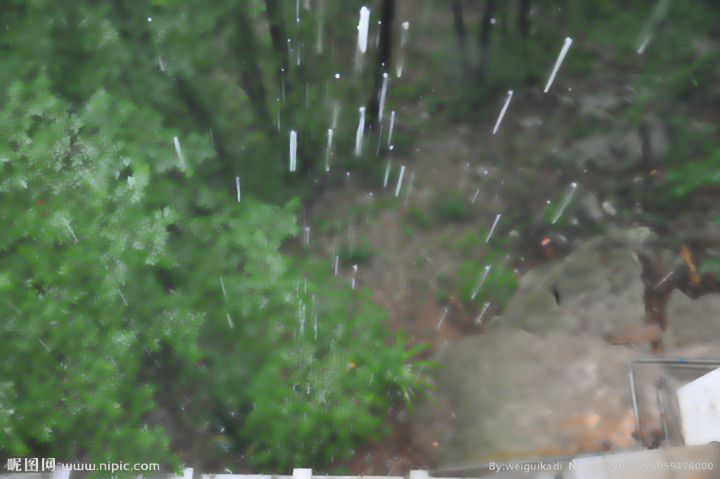}}
	\hspace{-1.9mm}	
	\subfigure{
		\includegraphics[width=2.9cm]{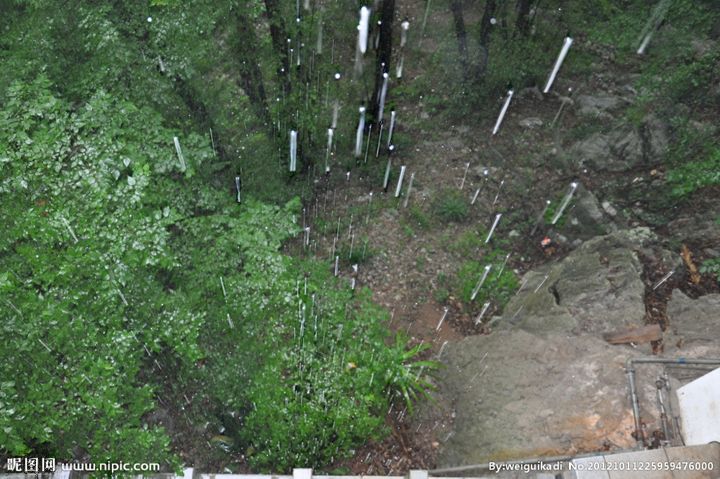}}
	\hspace{-1.9mm}	
	\subfigure{
		\includegraphics[width=2.9cm]{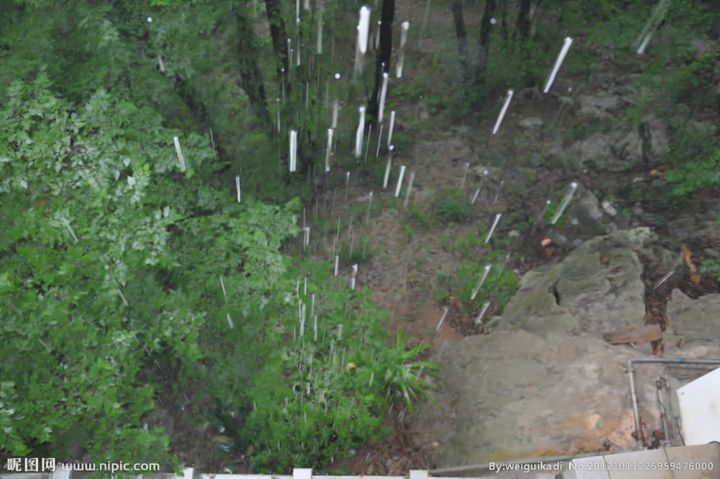}}
	\hspace{-1.9mm}	
	\subfigure{
		\includegraphics[width=2.9cm]{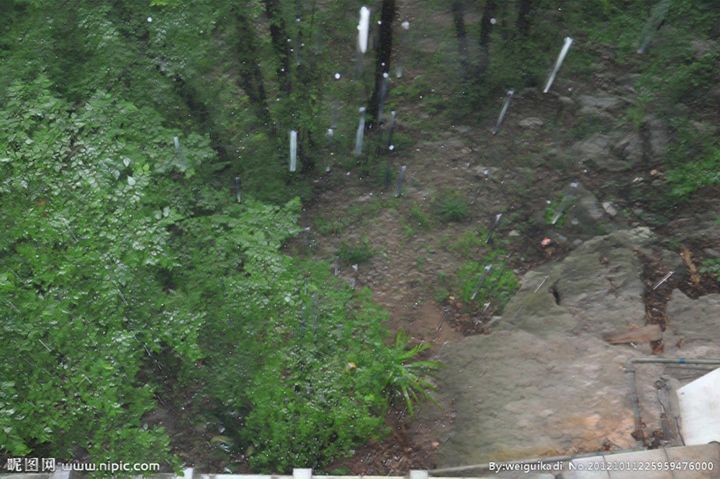}}
	\hspace{-1.9mm}	
	\subfigure{
		\includegraphics[width=2.9cm]{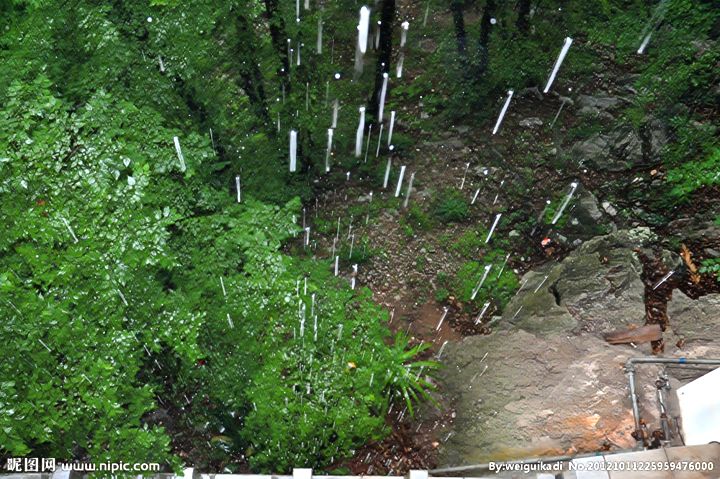}}
	\\ \vspace{-2mm}
	\subfigure{
		\includegraphics[width=2.9cm]{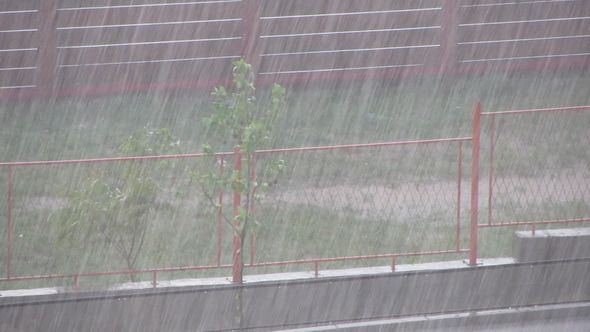}}
	\hspace{-1.9mm}	
	\subfigure{
		\includegraphics[width=2.9cm]{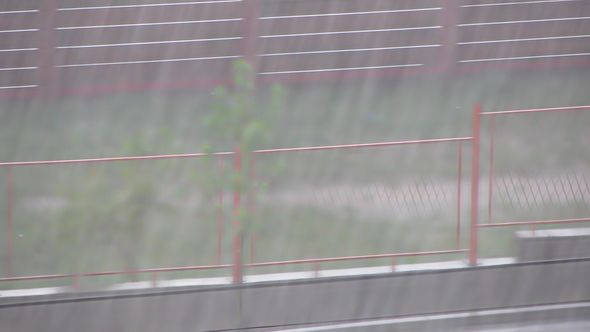}}
	\hspace{-1.9mm}	
	\subfigure{
		\includegraphics[width=2.9cm]{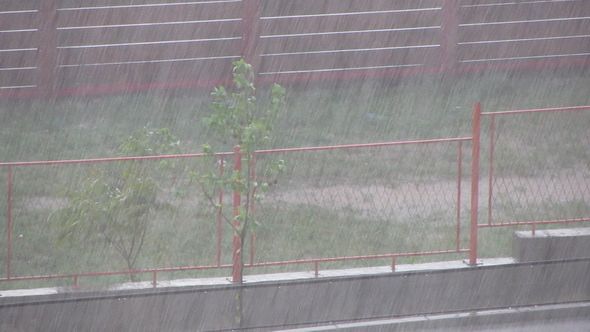}}
	\hspace{-1.9mm}	
	\subfigure{
		\includegraphics[width=2.9cm]{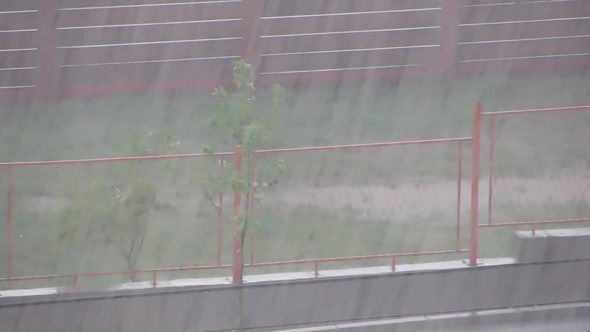}}
	\hspace{-1.9mm}	
	\subfigure{
		\includegraphics[width=2.9cm]{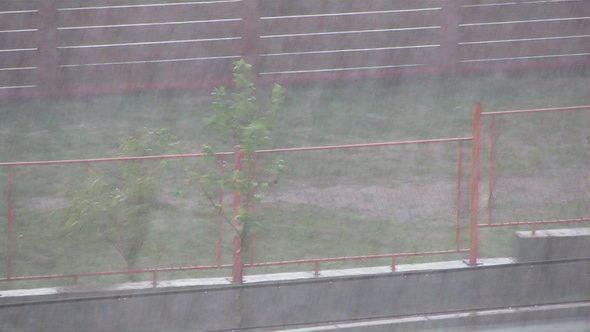}}
	\hspace{-1.9mm}	
	\subfigure{
		\includegraphics[width=2.9cm]{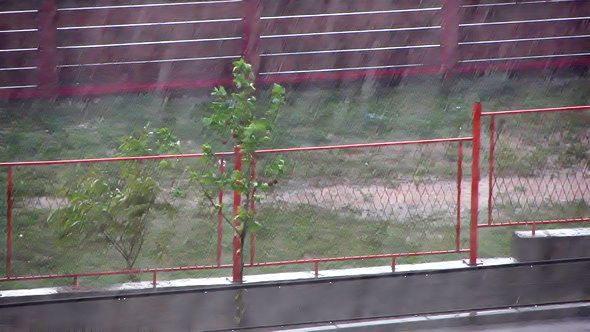}}
	\\ \vspace{-2mm}
	\subfigure{
		\includegraphics[width=2.9cm]{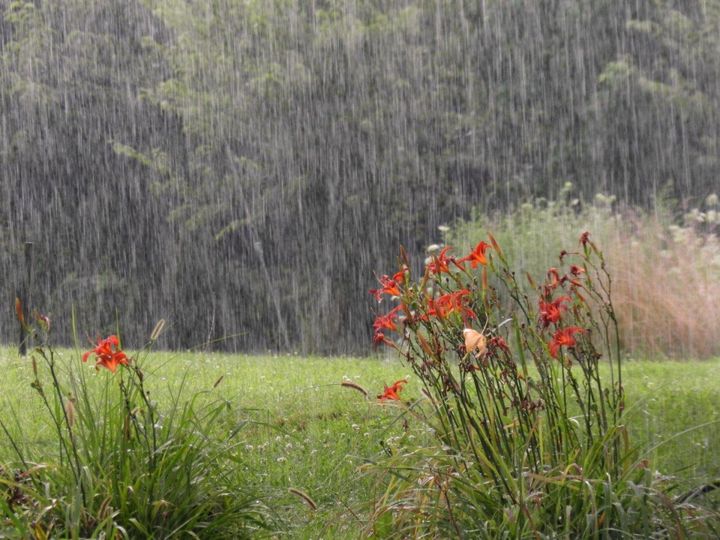}}
	\hspace{-1.9mm}
	\subfigure{
		\includegraphics[width=2.9cm]{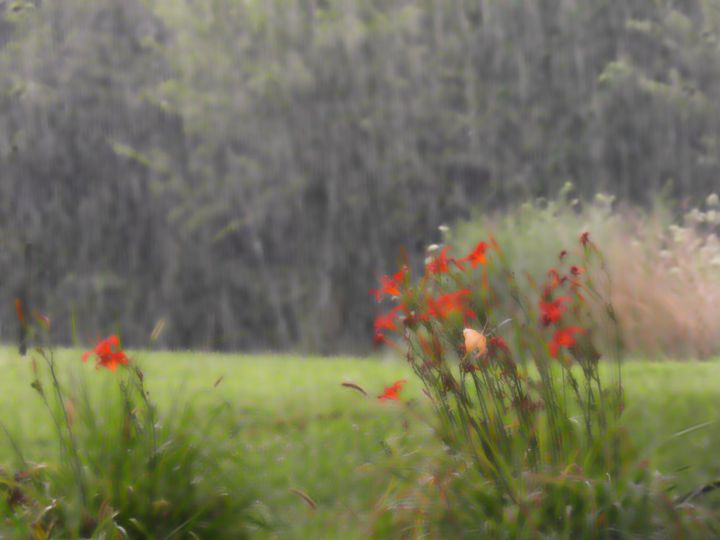}}
	\hspace{-1.9mm}	
	\subfigure{
		\includegraphics[width=2.9cm]{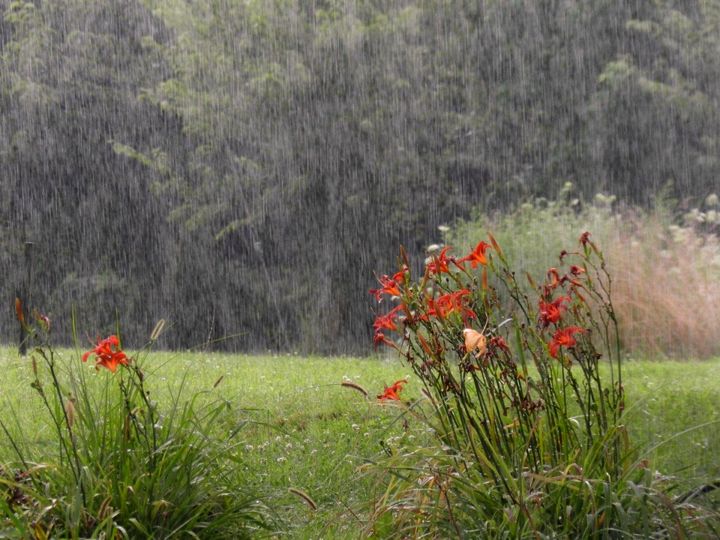}}
	\hspace{-1.9mm}	
	\subfigure{
		\includegraphics[width=2.9cm]{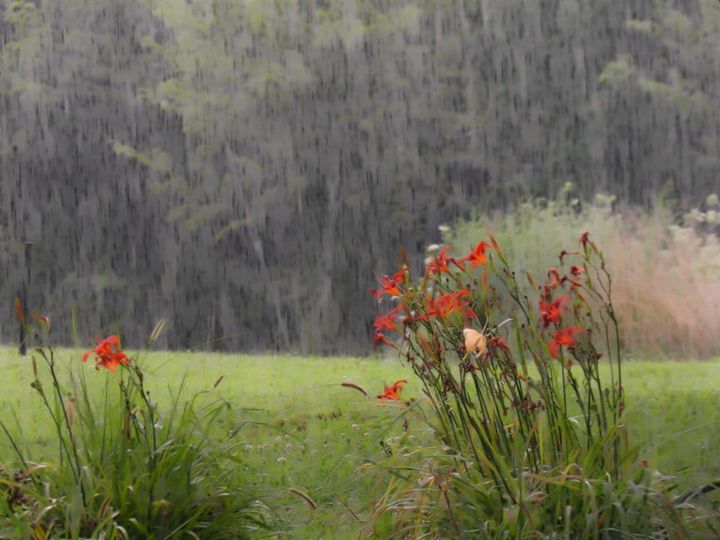}}
	\hspace{-1.9mm}	
	\subfigure{
		\includegraphics[width=2.9cm]{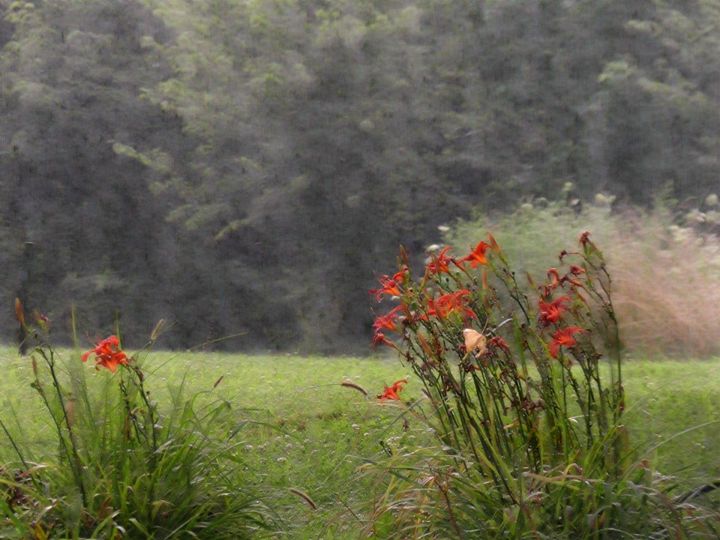}}
	\hspace{-1.9mm}	
	\subfigure{
		\includegraphics[width=2.9cm]{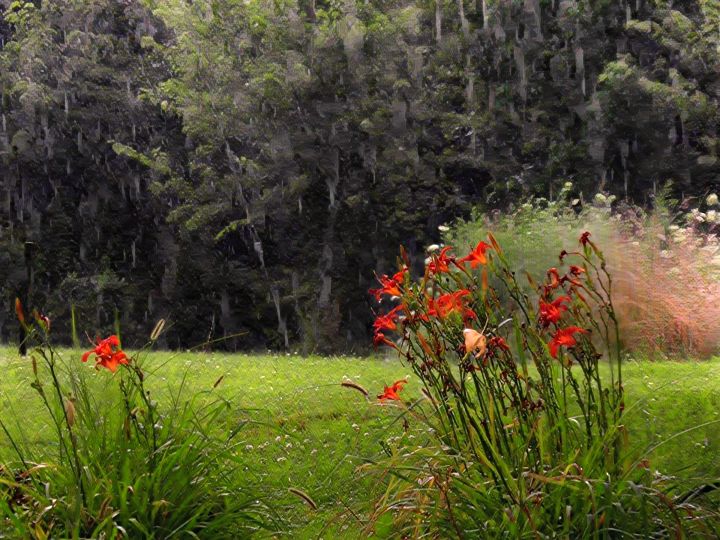}}	
	\\ \vspace{-2mm}	
	\setcounter{subfigure}{0}
	\subfigure[Input]{
		\includegraphics[width=2.9cm]{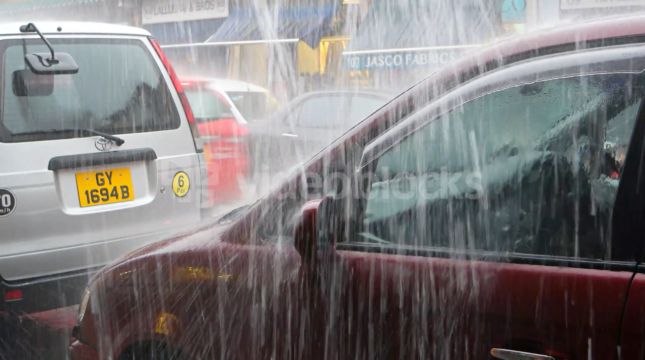}}
	\hspace{-1.9mm}	
	\subfigure[ID]{
		\includegraphics[width=2.9cm]{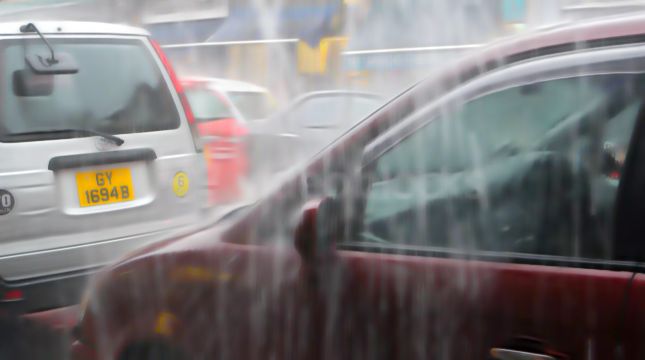}}
	\hspace{-1.9mm}	
	\subfigure[DSC]{
		\includegraphics[width=2.9cm]{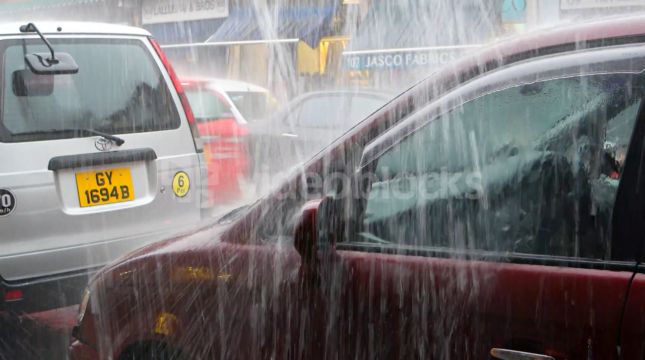}}
	\hspace{-1.9mm}	
	\subfigure[LP]{
		\includegraphics[width=2.9cm]{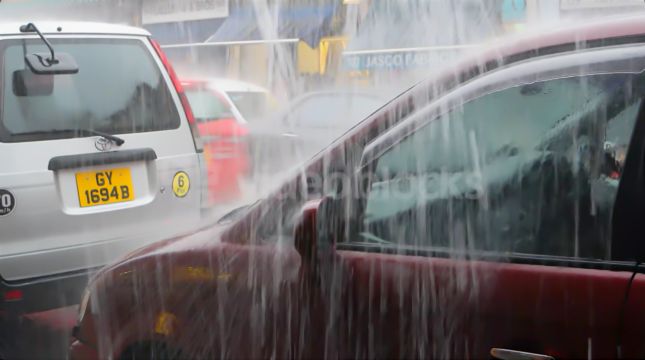}}
	\hspace{-1.9mm}	
	\subfigure[DetailNet]{
		\includegraphics[width=2.9cm]{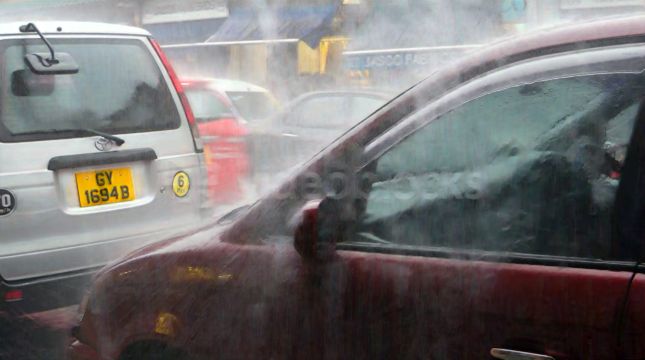}}
	\hspace{-1.9mm}	
	\subfigure[HeavyRainRemoval]{
		\includegraphics[width=2.9cm]{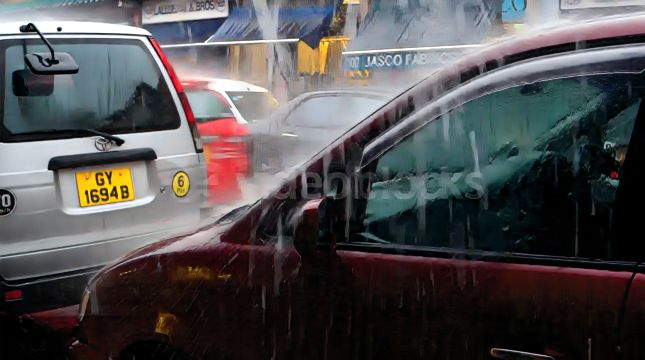}}
	\\ \vspace{-2mm}		
	\subfigure{
		\includegraphics[width=2.9cm]{Images/imgs1/1.jpg}} 
	\hspace{-1.9mm}	
	\subfigure{
		\includegraphics[width=2.9cm]{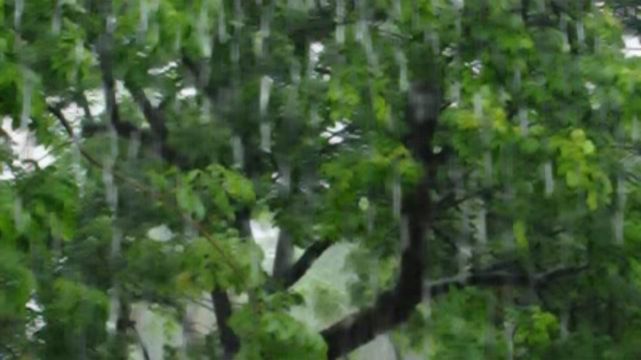}} 
	\hspace{-1.9mm}
	\subfigure{
		\includegraphics[width=2.9cm]{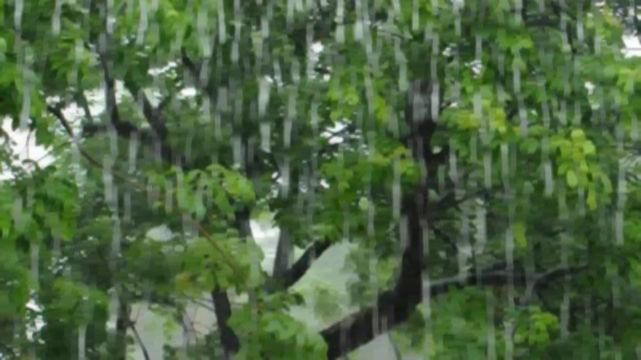}}
	\hspace{-1.9mm}	
	\subfigure{
		\includegraphics[width=2.9cm]{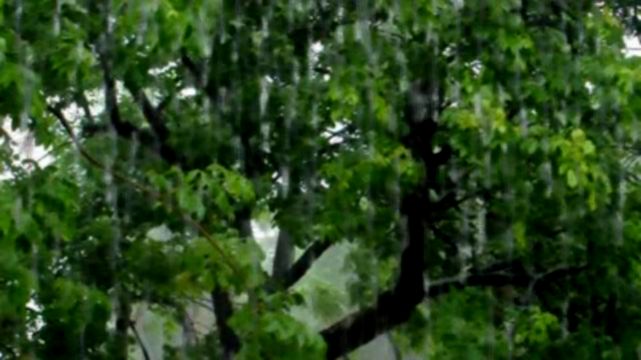}}
	\hspace{-1.9mm}	
	\subfigure{
		\includegraphics[width=2.9cm]{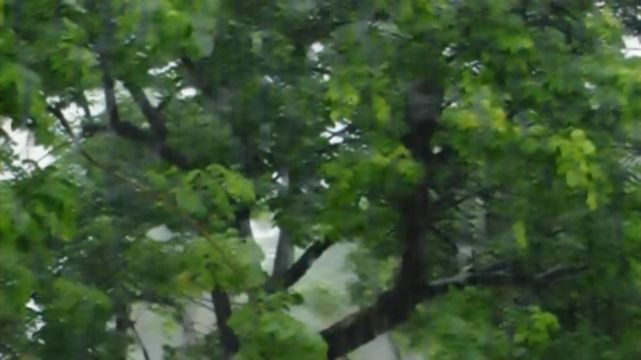}}
	\hspace{-1.9mm}
	\subfigure{
		\includegraphics[width=2.9cm]{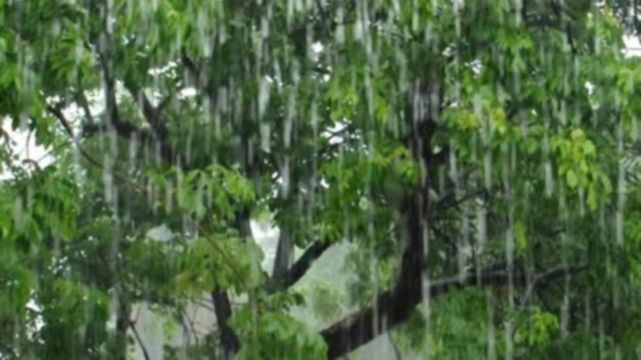}}	
	\\ \vspace{-2mm}
	\subfigure{
		\includegraphics[width=2.9cm]{Images/imgs2/1.jpg}} 
	\hspace{-1.9mm}	
	\subfigure{
		\includegraphics[width=2.9cm]{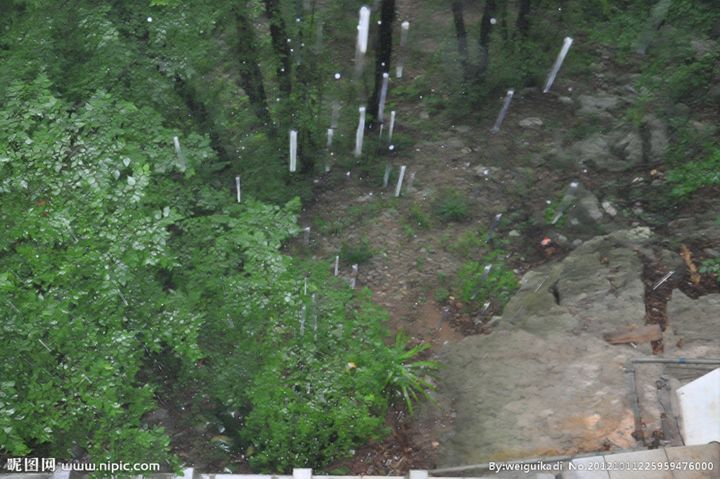}} 
	\hspace{-1.9mm}
	\subfigure{
		\includegraphics[width=2.9cm]{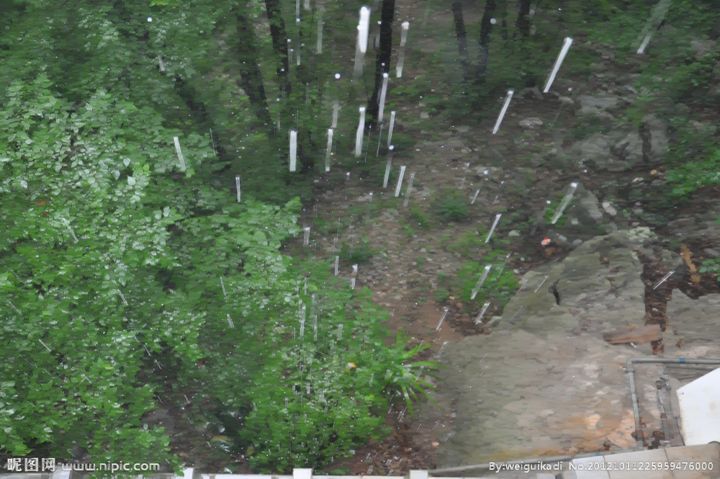}}
	\hspace{-1.9mm}	
	\subfigure{
		\includegraphics[width=2.9cm]{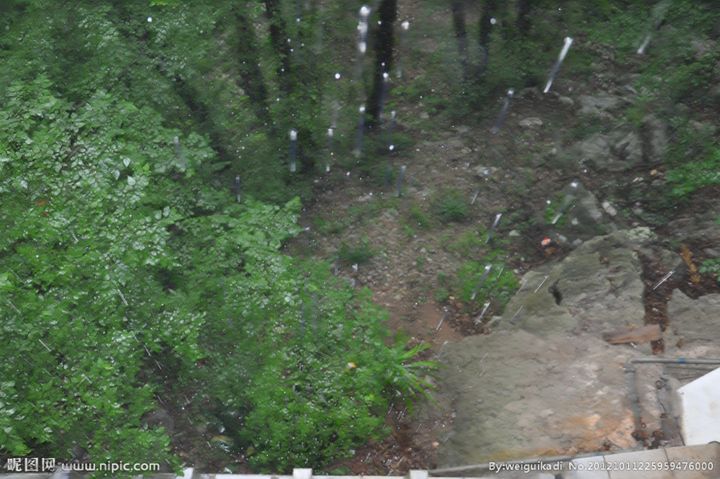}}
	\hspace{-1.9mm}	
	\subfigure{
		\includegraphics[width=2.9cm]{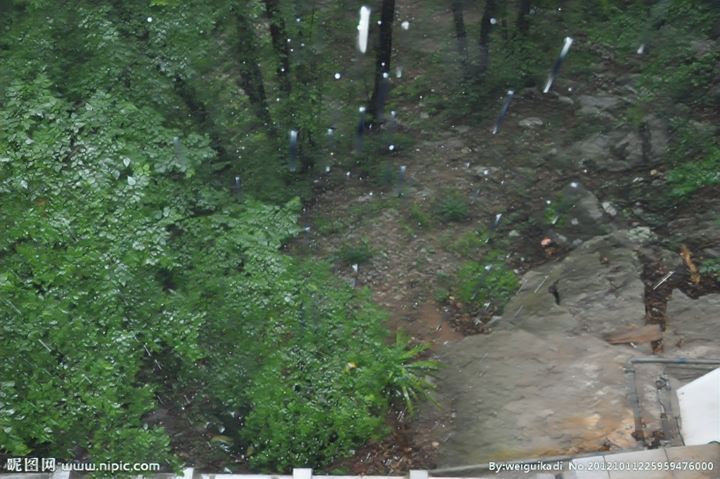}}
	\hspace{-1.9mm}
	\subfigure{
		\includegraphics[width=2.9cm]{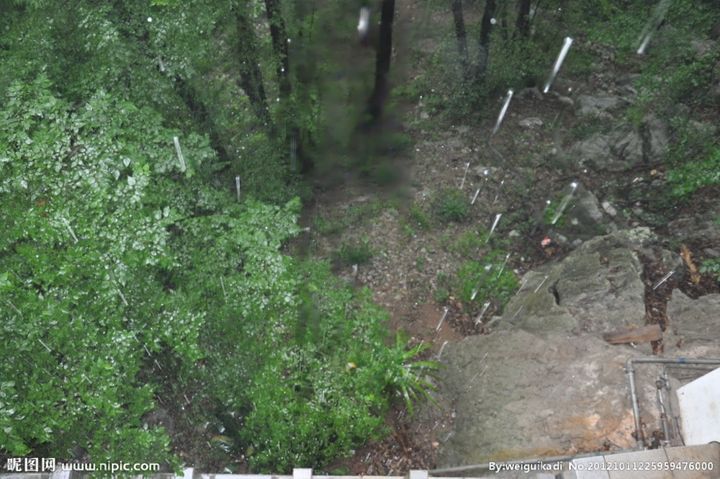}}
	\\ \vspace{-2mm}
	\subfigure{
		\includegraphics[width=2.9cm]{Images/imgs3/1.jpg}} 
	\hspace{-1.9mm}
	\subfigure{
		\includegraphics[width=2.9cm]{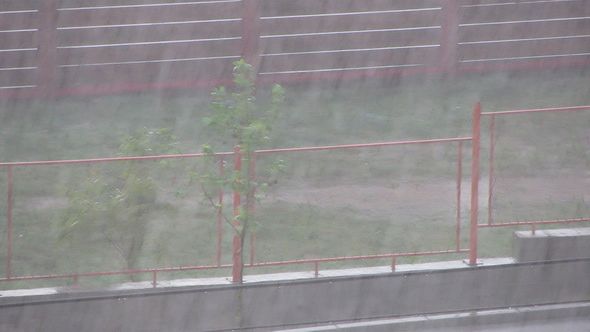}} 
	\hspace{-1.9mm}
	\subfigure{
		\includegraphics[width=2.9cm]{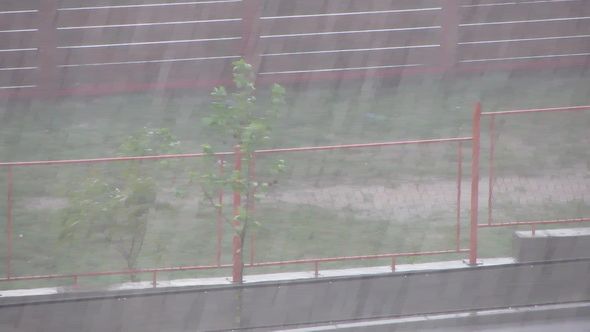}}
	\hspace{-1.9mm}	
	\subfigure{
		\includegraphics[width=2.9cm]{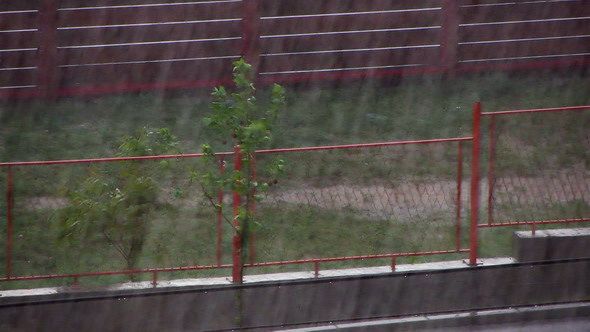}}
	\hspace{-1.9mm}	
	\subfigure{
		\includegraphics[width=2.9cm]{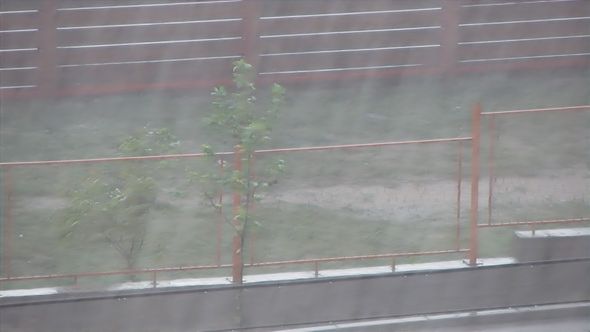}}
	\hspace{-1.9mm}
	\subfigure{
		\includegraphics[width=2.9cm]{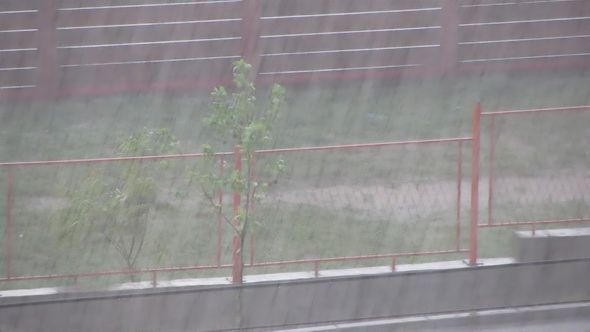}}
	\\ \vspace{-2mm}
	\subfigure{
		\includegraphics[width=2.9cm]{Images/imgs4/1.jpg}} 
	\hspace{-1.9mm}			
	\subfigure{
		\includegraphics[width=2.9cm]{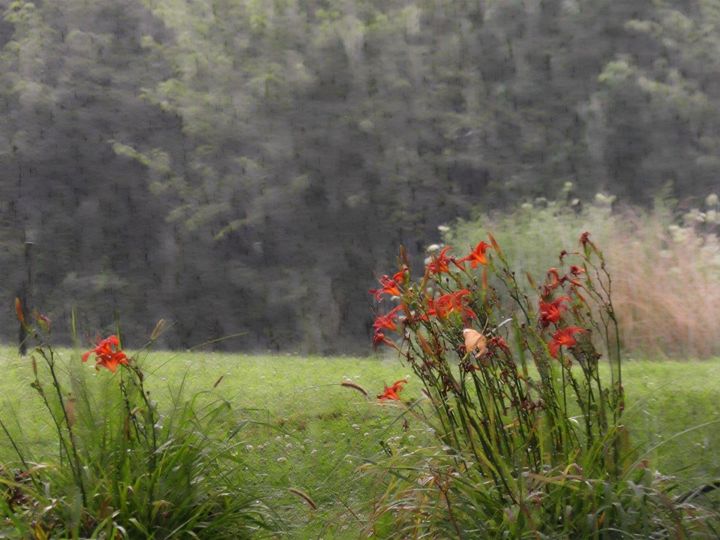}} 
	\hspace{-1.9mm}
	\subfigure{
		\includegraphics[width=2.9cm]{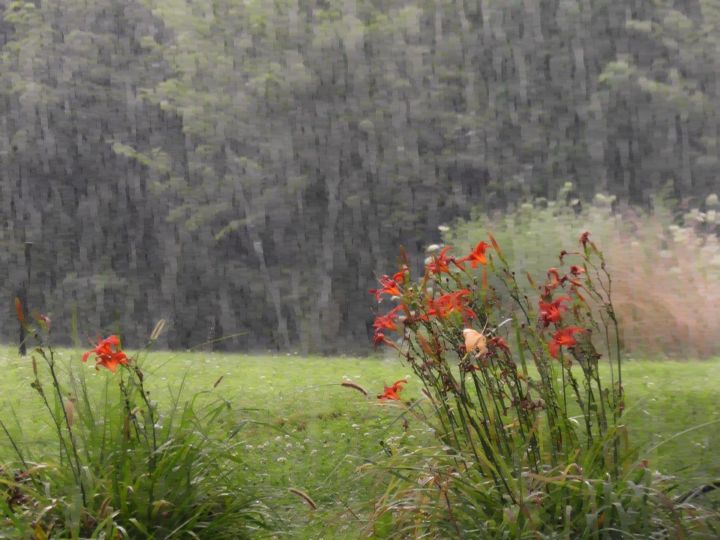}}
	\hspace{-1.9mm}	
	\subfigure{
		\includegraphics[width=2.9cm]{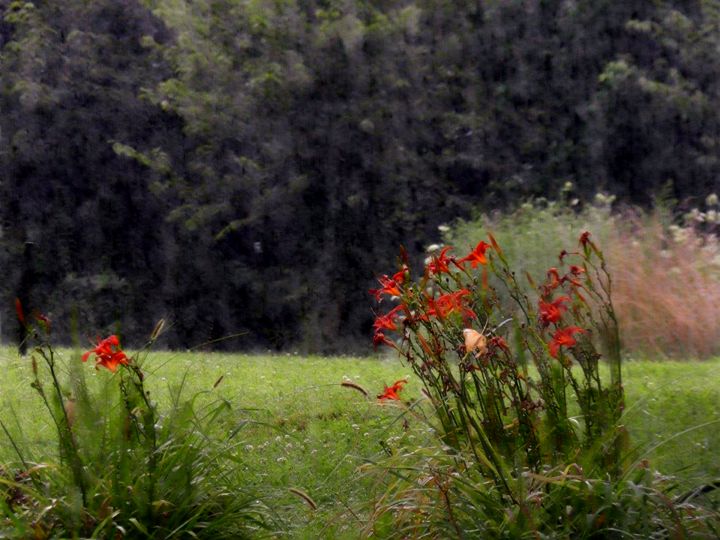}}
	\hspace{-1.9mm}	
	\subfigure{
		\includegraphics[width=2.9cm]{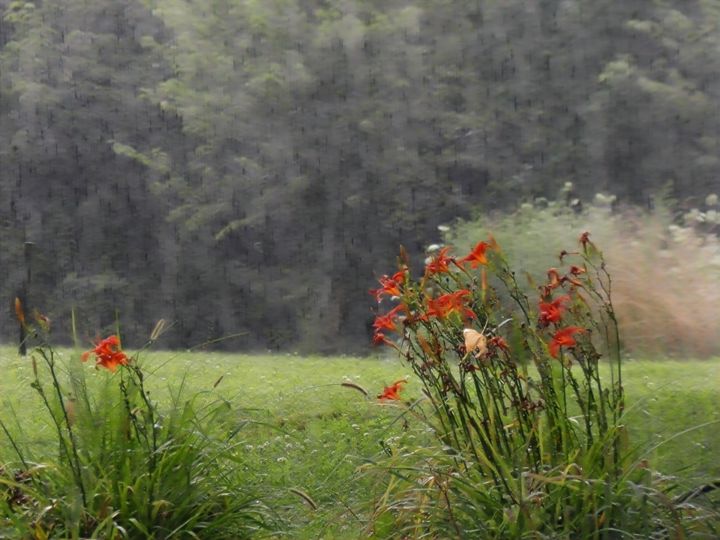}}
	\hspace{-1.9mm}
	\subfigure{
		\includegraphics[width=2.9cm]{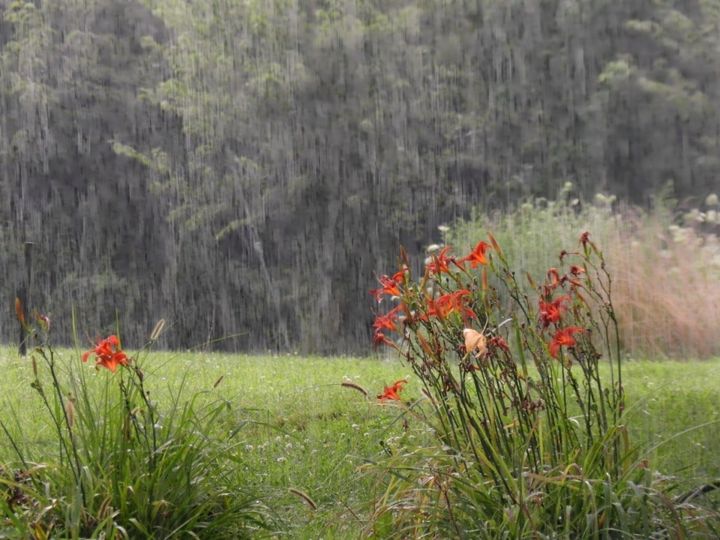}}	
	\\ \vspace{-2mm}
	\setcounter{subfigure}{6}
	\subfigure[DID-MDN]{
		\includegraphics[width=2.9cm]{Images/imgs5/1.jpg}} 
	\hspace{-1.9mm}		
	\subfigure[RESCAN]{
		\includegraphics[width=2.9cm]{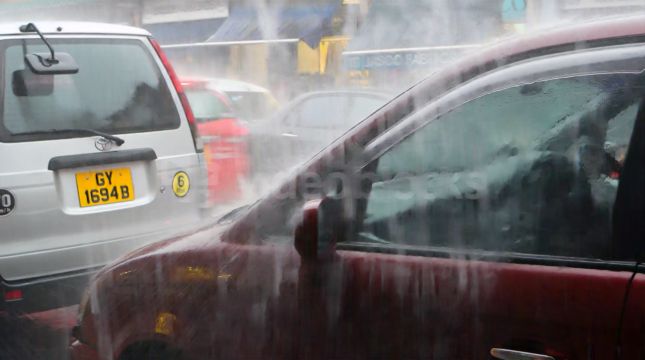}} 
	\hspace{-1.9mm}
	\subfigure[JCAS]{
		\includegraphics[width=2.9cm]{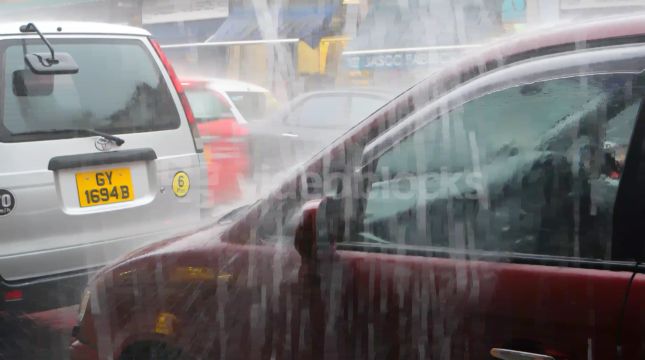}}
	\hspace{-1.9mm}	
	\subfigure[JORDER-E]{
		\includegraphics[width=2.9cm]{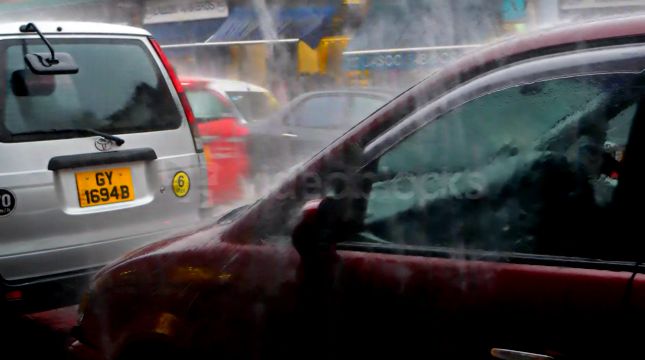}}
	\hspace{-1.9mm}	
	\subfigure[PReNet]{
		\includegraphics[width=2.9cm]{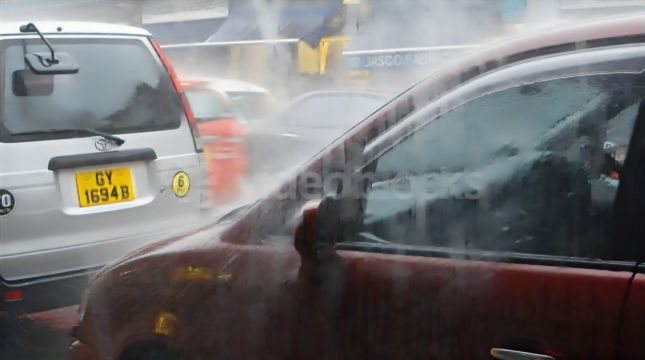}}
	\hspace{-1.9mm}
	\subfigure[SPANet]{
		\includegraphics[width=2.9cm]{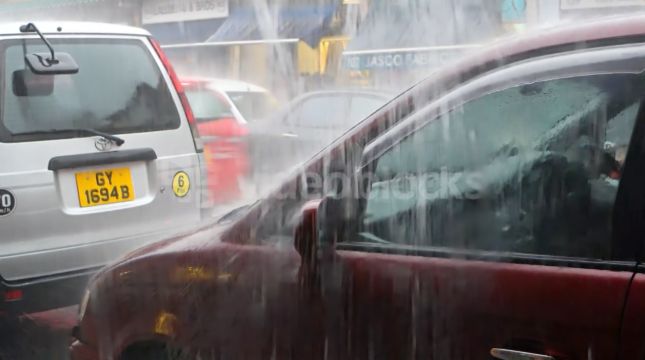}}	
	\\ 
	\caption{Visual results of different methods. 1\textit{st} and 2\textit{nd} panels: rain images with large rain streaks. 3\textit{th} and 4\textit{th} panels: rain images including rain accumulation. 5\textit{th} panel: rain images including both large rain streaks and accumulation.
	}
	\vspace{-5mm}
	\label{fig:sub_results}
\end{figure*}

We also show the visual results of different methods in Fig.~\ref{fig:sub_results}.
The input images shown in the figure are diversified and difficult to be handled, including large rain streaks and dense rain accumulation.
The top two panels clearly show that, JORDER-E (Fig.~\ref{fig:rain_results}~(j)) and PReNet (Fig.~\ref{fig:rain_results}~(k)) are better at handling large rain streaks.
JORDER-E (Fig.~\ref{fig:rain_results}~(j)) and HeavyRainRemoval (Fig.~\ref{fig:rain_results}~(f)) achieve better results in removing rain accumulation and enhancing the visibility from the bottom three panels.

We also use qualitative assessment metrics for performance comparison of different methods by evaluating the consistency between the results of deraining methods and the subjective results by Mean Opinion Score (MOS)\footnote{\wh{More information about the subjective evaluation dataset, results and the evaluation website can be found at: https://flyywh.github.io/Single\_rain\_removal\_survey/.}}.
There are 20 rain images for the evaluation.
These images are processed by the methods and their results are evaluated by human annotators.
40 participants are invited in the subjective experiment.
Each of them is required to provide subjective results for 550 image pairs.

\begin{figure}
	\flushleft
	\includegraphics[width=8cm]{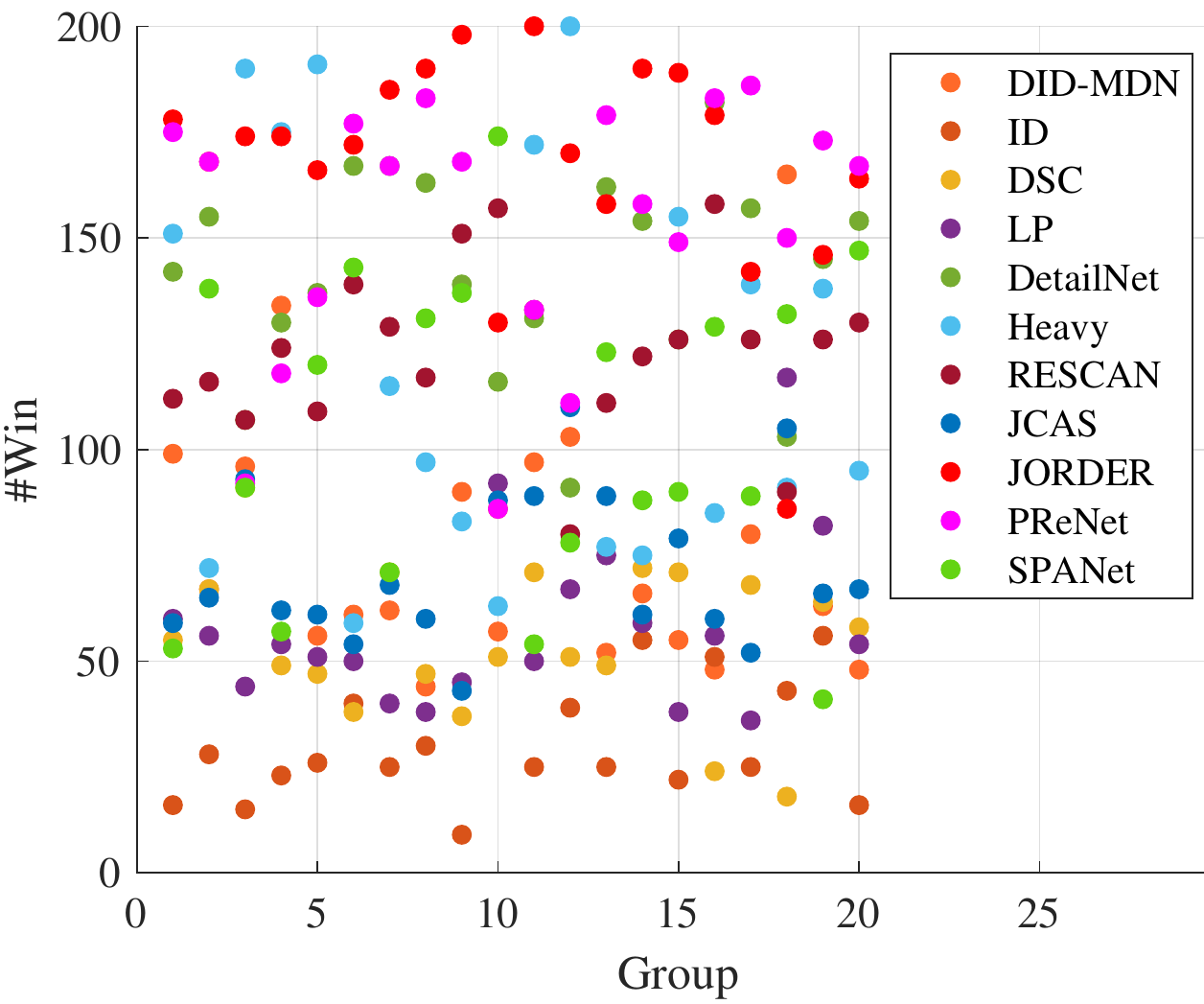}
	\vspace{-3mm}	
	\caption{Visualization of all paired comparisons. 
	The horizontal axis denotes the comparison group ID while the vertical axis denotes the winning time in the comparison.}
	\vspace{-3mm}
	\label{fig:visual_all}
\end{figure}

The comparison results are visualized in Fig.~\ref{fig:visual_all}.
Based on the compared pairs, we further fit a Bradley-Terry~\cite{Bradley} model to estimate the MOS score for each method so that they can be ranked.
We can infer the MOS score for each input sample, and then combine the results of different samples via geometric mean, which is denoted as the \wh{\textit{average MOS}} in Table~\ref{tab:nr_result}.
We can also directly infer the MOS score with the accumulated ranking results of all samples, which is denoted as \wh{\textit{MOS}} value in Table~\ref{tab:nr_result}.
%

In general, the paper published in 2019 are on average superior to previous methods on the dataset. 
However, the superiority of the qualitative comparison is not the same as that of the quantitative one,
which reflects the disagreements between optimizing the quantitative metrics on the synthesized data and achieving better visual quality on real images. This is due to the  domain gap between the real rain images and synthesized data.

We also observe that, all non-reference metrics are not in agreement with MOS and the qualitative values. 
We calculate Spearman rankorder correlation coefficient (SROCC),
Kendallrank-order correlation coefficient (KROCC),
and Pearson linear correlation coefficient (PLCC)
in Table~\ref{tab:evaluation}, where large absolute values denote that the metric can obtain more consistent results with respect to human perception.
One can see that the values for the best result are only 0.2216, 0.1473, 0.1864 for SROCC, KROCC, and PLCC, respectively.

We conclude that all the existing metrics are not suitable to measure the performance of deraining, and thus there is great potential for future works on the deraining performance evaluation.

\begin{table}[t]
	\centering
	\caption{The evaluation results of all quality assessment models.}
	\label{tab:evaluation}	
	\begin{tabular}{c|ccc}
		\hline
		Methods   & SROCC   & KROCC   & PLCC   \\
		\hline
		\hline
		NIQE      & 0.0780  & 0.0461  & 0.0700 \\
		PIQE      & 0.2118  & 0.1437  & 0.1215 \\
		BRISQUE   & 0.1896  & 0.1297  & 0.1508 \\
		ILNIQE    & 0.0778  & 0.0508  & 0.1458 \\
		SSEQ      & 0.2216  & 0.1473  & 0.1257 \\
		SR-Metric & 0.1132  & 0.0760  & 0.1129 \\
		ENIQA     & 0.1333  & 0.0932  & 0.1487 \\
		BIQAA     & 0.1365  & 0.0927  & 0.1383 \\
		BIQI      & 0.2001  & 0.1299  & 0.1558 \\
		BLIINDS2  & 0.1705  & 0.1208  & 0.1752 \\
		FRISQUE   & 0.2083  & 0.1407  & 0.1864 \\
		\hline
	\end{tabular}
\end{table}

\begin{table*}[htbp]
	\centering
	\caption{The model complexity and running time of different methods. The size of the input image is $512\times512$. C and G denote CPU and GPU, respectively.}
	\label{tab:complextiy}
	\begin{tabular}{c|ccccccc}
		\hline
		Baseline & ID & DSC & LP & UGSM & JCAS & DetailNet & DID-MDN  \\ \hline \hline
		Time (Seconds)  & 283.69 & 398.66 & 1177.3 & 2.51 & 188.28 & 0.61 & 0.53  \\ 
		GPU/GPU  & (C) & (C) & (C) & (C) & (C) & (G) & (G)  \\ 
		Para.    & - & - & - & - & - & 57,369 & 372,839  \\ \hline
		Baseline & JORDER-E  & RESCAN  & ID-CGAN  & SPANet & URML & HeavyRainRemoval & PReNet \\ \hline \hline
		Time (Seconds)  & 0.13 & 0.61    & 0.50    & 1.72 & 2.02 &  0.73 & 0.11 \\ 
		GPU/GPU  & (G) & (G)     & (G)     & (G) & (G) & (G) & (G) \\ 
		Para.    & 4,169,024  & 149,823 & 263,686 & 283,716 & 984,356 & 40,627,038 & 168,963 \\ \hline		
	\end{tabular}
\end{table*}

\subsection{Computational Complexity}

Table~\ref{tab:complextiy} compares the runtime of different state-of-the-art methods. All sparse coding based methods are implemented in MATLAB and tested on a CPU, following the original setting of  all the released codes, while other methods are accelerated by a GPU.
ID-CGAN is implemented in Torch7. 
The rest is implemented in Pytorch.
One can observe that JORDER-E, HeavyRainRemoval, and URML employ many more parameters than other methods.
The comparison results on both performance and parameter number show that 
PReNet is an impressive method quantitatively and qualitatively, while keeping a light-weighted framework.

\section{Future Directions}

\subsection{Integration of Physics Model and Real Images}

Many existing learning-based methods rely on synthetic rain images to train the networks, since to obtain paired rain images and their exact clean ground-truths is intractable. While such a training scheme shows some degree of success, to improve the performance, we need to incorporate both real rain images the training process; otherwise, the network will never been exposed to the real rain images, impeding the network's effectiveness in the testing stage.
Incorporating real rain images, however, can pose problems, because to obtain the paired clean background images is intractable. Consequently, there is no loss for a network to learn. To address this problem, we may rely on physics-based constraints.
An attempt in~\cite{Lil_2019_CVPR} has shown the feasibility of this direction. Specifically, it combines the power of a physics model and a generative adversarial network, which can accept unpaired ground-truths.
In the future, more works are expected in this direction to make efforts to combine of physical models and real rain images.

\subsection{Rain Modeling}
The current synthetic rain models can only cover limited types of rain streaks, \textit{e.g.} a range of scales, shapes, directions, \textit{etc.} 
However, in practice, the appearance of rain streaks is diverse, due to many different factors that can influence rain conditions, \textit{e.g.} 3D environments, distances, wind directions/speed, \textit{etc.}
Currently, when the distributions of captured rain streaks are different from the synthetic images in the training, the methods tend to fail to remove rain properly.
The studies of \cite{Wei_2019_CVPR,UD_GAN} attempt to model the rain appearance via the generation model and unpaired learning.
However, observing their generated rain images, one can visibly see that they are not as diverse as real rain can be, and they are visibly not real enough. The latter can also cause problems, since it means there are significant gaps between synthetic and real rain images.

\subsection{Evaluation Methodology}

With a rapid growth of works on rain removal, 
it is still challenging to measure whether a method is sufficiently effective.
As shown in Sec.~\ref{sec:subjective}, existing quality assessment methods are still far from capturing real visual perception of human.
Thus, there is a potential direction which the community can pay more attention to.
The quality assessment of rain removal methods can be considered from two aspects.
First, for human vision, the metric should be designed to model the typical distortions caused by rain and deraining methods, and to describe the human preferences to different deraining results.
Second, for machine vision, we could consider the performance of high-level vision tasks in rain conditions.
The MPID dataset makes the preliminary attempt by constructing task-driven evaluation sets for traffic detection.
In the future, we hope that more large-scale task-driven evaluation sets with more applications in more diverse rain conditions.

\subsection{More Related Tasks and Real Applications}

When existing deraining methods are applied to real applications, there are a few factors  that should be considered.
First, the runtime of the method.  Current methods are far from the requirement of real-time processing (30 fps).
How to accelerate existing methods is a future challenge.
Second, real rain images usually contain more complicated visual degradation. 
For example, the surveillance videos are compressed and also include compression distortion, \textit{e.g.} blocking artifacts. Effective deraining methods also need to take care of these issues.
Third, there are scenarios where composite degradation might be involved, \textit{e.g.}
night-time rain conditions, mixture of raindrop and rain streak, \textit{etc}.
It will be interesting to detect the degradation types and handle them in a unified framework adaptively.

\section{Concluding Remarks}

We have surveyed single-image deraining methods based on model-based and data-driven approaches. We discussed the rain models, the challenges of single-image deraining, and the basic ideas of the model-based and data-driven methods.
In our discussion, the model-based methods are categorized further into: layer decomposition, sparse coding and GMMs; and, data driven based methods are grouped into: deep CNN, generative adversarial network, and semi/unsupervised learning methods.
We also learned that data-driven methods generally perform better than the model-based methods.
However, there are still a few open problems, particularly in the data-driven approach. Problems such as fusing physics models and real-rain images, more accurate rain models, evaluation methodology, and real applications of deraining still need further developments.

\ifCLASSOPTIONcaptionsoff
\newpage
\fi

\renewcommand{\baselinestretch}{1}

\bibliographystyle{IEEEtran}

\end{document}